\documentclass[12pt]{article}

\usepackage{graphics, color}
\usepackage{graphicx}
\usepackage{amssymb}
\usepackage{amsmath}
\pdfoutput=1

\newcommand{\sect}[1]{\section{#1}\setcounter{equation}{0}}

\def\gsim{\, \rlap{$>$}{\lower 1.1ex\hbox{$\sim$}}\,}
\def\lsim{\, \rlap{$<$}{\lower 1.1ex\hbox{$\sim$}}\,}

\def\R{{{\cal R}}}

\def\Op{{\mathcal{O}}}
\def\V{{\mathcal{V}}}

\def\kUV{\kappa}
\def\kIR{\kappa_{\rm IR}}

 \newcommand{\be}{\begin{equation}}
\newcommand{\ee}{\end{equation}}
 \newcommand{\bal}{\begin{align}}
 \newcommand{\eal}{\end{align}}
\newcommand{\ben}{\begin{equation*}}
\newcommand{\een}{\end{equation*}}
\newcommand{\bea}{\begin{eqnarray}}
\newcommand{\eea}{\end{eqnarray}}
\newcommand{\bean}{\begin{eqnarray*}}
\newcommand{\eean}{\end{eqnarray*}}
\newcommand{\bes}{\begin{subequations}}
\newcommand{\ees}{\end{subequations}}

\def\wAdS{{\widetilde{AdS}}}
\def\wDelta{{\widetilde{\Delta}}}
\def\walpha{{\widetilde{\alpha}}}

\usepackage{epsfig}

\begin{document}


\begin{titlepage}
\bigskip
\bigskip\bigskip\bigskip
\centerline{\Large Holographic quantum criticality from multi-trace deformations}
\bigskip\bigskip\bigskip
\bigskip\bigskip\bigskip

\centerline{{\bf Thomas Faulkner}}
\medskip
\centerline{\em Kavli Institute for Theoretical Physics}
\centerline{\em University of California}
\centerline{\em Santa Barbara, CA 93106-4030}\bigskip
\medskip
\centerline{{\bf Gary T. Horowitz}}
\medskip
\centerline{and}
\medskip
\centerline{{\bf Matthew M. Roberts}}
\medskip
\centerline{\em Department of Physics}
\centerline{\em University of California}
\centerline{\em Santa Barbara, CA 93106-4030}\bigskip

\bigskip
\bigskip\bigskip


\begin{abstract}
We explore the consequences of multi-trace deformations in applications of gauge-gravity duality to condensed matter physics. We find that they introduce a powerful new ``knob" that can implement spontaneous symmetry breaking, and can be used to construct a new type of holographic superconductor. This knob can be tuned to drive the critical temperature to zero, leading to a new quantum critical point. We calculate nontrivial critical exponents, and show that fluctuations of the order parameter are `locally' quantum critical in the disordered phase. Most notably the dynamical critical exponent is determined by the dimension of an operator at the critical point. We argue that the results are robust against quantum corrections and discuss various generalizations.

\end{abstract}
\end{titlepage}
\baselineskip = 17pt
\tableofcontents
\setcounter{footnote}{0}

\sect{Introduction}

Over the past couple of years, gauge/gravity duality has been applied to a number of problems in condensed matter physics (for reviews see \cite{Hartnoll:2009sz,McGreevy:2009xe,Sachdev:2010ch}).  An important feature of some condensed matter systems is the existence of quantum
critical points, marking continuous phase transitions at zero temperature. 
One goal of the present work is to introduce and study a new mechanism for generating quantum critical points in the context of gauge/gravity duality. We will see that the behavior near the critical points is described by nontrivial critical exponents and goes beyond the
usual Landau-Ginzburg-Wilson description of phase transitions at zero temperature.

A second goal is to introduce a new type of holographic superconductor. The key ingredient in constructing  a gravitational dual of a superconductor is to find an instability which breaks a $U(1)$ symmetry at low temperature and causes a condensate to form. Previous constructions have started with a charged anti de Sitter (AdS) black hole which has such an instability when coupled, e.g.,  to a charged scalar field \cite{Gubser:2008px,Hartnoll:2008vx,Hartnoll:2008kx}. We will show that there is another source of instability which applies even for Schwarzschild AdS black holes. So these new superconductors can exist even at zero chemical potential and  no net charge density.

Both of these goals are achieved by adding a multi-trace operator to the dual field theory action\footnote{For another recent discussion of multi-trace operators in gauge/gravity duality, see \cite{Vecchi:2010dd}.}. For example, given a  (single trace) scalar operator $\mathcal{O}$ of dimension $\Delta_- <  3/2$ in a $2+1$ dimensional field theory (which will be our main focus) one can modify the action 
\begin{equation}
\label{eq:dblUV}
S \rightarrow S -  \int d^3x\,  \bar{\kappa} \, \mathcal{O}^\dagger
\mathcal{O} 
\end{equation}
where for convenience later the coupling will be rescaled as $\bar{\kappa} = 2( 3- 2\Delta_-)  \kappa$.
Since this  is a relevant deformation, it is unnatural  to exclude such a term, and it has important consequences. If $\mathcal{O}$ is the operator dual to the bulk charged scalar field in conventional holographic superconductors, then adding this term (with $\kUV> 0$) makes it harder to form the condensate and lowers the critical temperature. We will see that in some cases $T_c$ vanishes at a finite value of $\kUV = \kappa _c$. This defines a new quantum critical point which we will study in detail. Since $T_c$ can be quite large at $\kUV = 0$, adding this double trace perturbation introduces a sensitive new knob for adjusting the critical temperature.

For $\kUV > \kappa_c$ (and nonzero chemical potential $\mu$), the ground state is described by the extremal Reissner-N\"ordstrom (RN) AdS black hole which has an emergent $AdS_2$  geometry in the IR. For $\kUV < \kappa_c$ there are various possible IR geometries depending on details of the bulk potential. However, as $\kUV$ approaches $\kappa_c$ from below, the bulk solution develops an intermediate $AdS_2$ geometry. It is this intermediate region which controls the behavior near the critical point. For example, we will show that the critical exponents do not take  mean field values, but are determined by the scaling dimension of  certain operators in the $0+1 $ dimensional CFT dual to this region. This is closely analogous to the way properties of holographic non-Fermi liquids \cite{Lee:2008xf,Liu:2009dm,Cubrovic:2009ye}  were described in terms of a dual $0+1 $ dimensional CFT \cite{Faulkner:2009wj,Faulkner:2010da}. 
In addition,  the instability for $\kUV < \kappa_c$ can be interpreted as turning on a double trace term with negative coefficient in the $0+1 $ dimensional CFT dual to this region. 

Let us contrast this with the usual argument for why the RN AdS black hole becomes unstable at low temperature in the presence of  a scalar field \cite{Denef:2009tp}. In the $AdS_2$ near horizon geometry of the $T=0$ solution, the scalar field has an effective mass $m_{\rm eff}$ which  depends on the original mass $m$ and charge $q$ of the scalar field. When this effective mass squared is below the Breitenlohner-Freedman (BF) bound \cite{BF82} for $AdS_2$, this near horizon region becomes unstable. Since $m^2$ is above the BF bound of the asymptotic $AdS_4$ geometry, the asymptotic region is stable, and the solution settles down to a black hole with scalar hair. It is now clear that this argument is sufficient but not necessary. It overlooks the possibility of instabilities with $m_{\rm eff}^2$ above the BF bound which are allowed due to modified boundary conditions for the scalar field. The boundary conditions may be modified due to the addition of a multi-trace deformation in the dual field theory \cite{Witten:2001ua,Berkooz:2002ug}, or simply due to alternative quantization of the bulk theory \cite{Klebanov:1999tb}.

It was widely believed that if one added a double trace term with $\kUV<0$, then the theory would not have a stable ground state. However, we have recently shown that this is not necessarily the case \cite{Faulkner:2010fh}.  For a large class of dual gravity  theories, there is still a stable ground state with $\langle\mathcal{O}\rangle \ne 0 $ when the boundary conditions correspond to $\kUV < 0$. As one increases the temperature, there is a second order phase transition to a state with  $\langle\mathcal{O}\rangle = 0 $.  This provides a new mechanism for spontaneously breaking a $U(1)$ symmetry and constructing novel holographic superconductors. This mechanism does not require a charged black hole and works for Schwarzschild AdS as well. In other words, one can set $\mu =0 $ and still break the $U(1)$ symmetry at low temperature. The critical temperature is now set by $\kUV$. We will discuss some properties of these novel holographic superconductors in section 2.

It is worth pointing out that the coupling $\kappa$, as the coefficient
of the square of the order parameter,  is the usual
tuning parameter in the context of Landau-Ginzburg theory. Also
if $\mathcal{O}$ is a gauge invariant trace of a fermion bilinear then
the double trace is a 4 fermion interaction, a natural interaction
to consider.

Although our discussion so far has focussed on the case where the operator $\mathcal{O}$ is charged, our results apply equally well when $\mathcal{O}$ is neutral. In this case, the ordered phase breaks a $Z_2$ symmetry. More generally, one can imagine different symmetry breaking scenarios where for example  $\mathcal{O}$ could be part of a triplet of operators forming a representation of $SU(2)$ which is spontaneously broken at low temperature. 
This is a particular attractive possibility as outlined in \cite{Iqbal:2010eh,Faulkner:2010tq}, since
in many condensed matter systems $SU(2)$ spin is a global symmetry
(ignoring spin orbit effects.) Including an exact global $SU(2)$ symmetry
then allows us to model magnetism in a holographic setup.
The triplet in which one embeds $\mathcal{O}$ can be interpreted as the
staggered order parameter associated with anti-ferromagnetic transitions.
Since the boundary theory has a global $SU(2)$ symmetry
the bulk will have an $SU(2)$ gauge symmetry distinct from the $U(1)$
electromagnetic charge. For the rest of this paper the $SU(2)$
gauge fields and triplet structure of the order parameter will not play a roll.

It is useful to bear this possibility in mind, in particular so we can
compare our results to quantum phase transitions 
in metallic systems, where anti-ferromagnetic
order plays an important role. The ``standard'' theory
of which was given in \cite{hertz,millis} 
is based on the renormalization group Landau-Ginzburg 
paradigm.
However experimental measurements (see for example \cite{vojtarev,sirev}, 
and references therein) of heavy
fermion systems with quantum critical points show
that the ``standard'' theory can break down,
as a result new theoretical methods are required. 
Subsequently several different methods were developed (see for example \cite{si1,si2,senthil1,senthil2}.)
One such method \cite{si1,si2} which is formally justified in a large $d$ expansion \cite{siedmft}
shows  local quantum critical behavior similar
to the new quantum critical point that we find. 
It will be useful to compare and contrast our results to those of the ÒstandardÓ theory and the 
other theoretical methods used in the study of quantum criticality in heavy
fermion systems. 

Since the finite density normal phase we consider is governed
by $AdS_2 \times R^2$ in the IR, at zero temperature the
theory has a finite entropy density. This has lead
many people to suggest that this state must not
be the true ground state, since otherwise one finds unnatural violations
of the third law of thermodynamics. Of course this may be natural in the context
of applications to heavy fermion system  where
superconductivity instabilities can be observed close to criticality. So
it may be that the state we work in is the correct one for a large range
of temperatures, but ultimately at low temperatures something else
takes over. Indeed as is discussed and extended in this
paper  $AdS_2$ has many possible forms of instability.
This however motivates us to attempt to extend our results
in various directions to directly address this problem. One extension
we consider is adding a magnetic field. We show
that while a magnetic field suppresses the superconducting
instability, it can enhance the neutral (anti-ferromagnetic) instability.
Another extension we consider is replacing $AdS_2 \times R^2$
with other possible IR geometries, such as a Lifshitz geometry
which does not have finite entropy density at zero temperature. 
We find our results are rather robust here.

The organization of the paper is as follows: In the next section
we show how double trace deformations
can induce spontaneous symmetry breaking and use this to construct a new type of holographic superconductor with zero net charge density.  We then extend this to the finite density
case and show that  the coefficient of the
double trace deformation provides a sensitive knob
by which one can tune the critical temperature $T_c$ to zero.
In section 3 we study the new quantum critical point that arises and analytically compute the nontrivial critical exponents. In section 4 we numerically
construct the backreacted geometries that correspond
to the ordered (condensed) phase away from the phase transition. We confirm the critical exponents near the critical point. In the discussion section, we summarize our results and discuss generalizations to magnetic fields and Lifshitz normal phases. The Appendices contain additional details.

\sect{Double trace deformations}

In this section we begin by emphasizing the simple under appreciated fact that double
trace deformations are useful for studying symmetry breaking
in gauge/gravity duality. We then note that this system provides a simple
holographic model for superconductivity with zero total charge density. With nonzero charge density, we show that double trace deformations introduce a new parameter by which one can tune the critical temperature of the superconductor. 

\subsection{Gravity setup and boundary conditions}

The theory we study is gravity in $3+1$ dimensions with a negative cosmological
constant, a $U(1)$ gauge field, and a scalar field $\Psi$ which may or may
not be charged under the $U(1)$ symmetry.  By general arguments
of gauge/gravity duality this theory is dual to a $CFT_{2+1}$ with
a conserved current operator $J^\mu$ and a scalar operator $\mathcal{O}$.

The action is that of the Einstein-Abelian Higgs model with a negative cosmological constant, where we parameterize the phase and modulus of the charged scalar as $\Psi = \psi e^{i\theta}$, following e.g. \cite{Aprile:2010yb}

\begin{equation}
\label{action}
S =  \int d^4x \sqrt{-g} \left( R  - \frac{1}{4} G(\psi) F^2 - (\nabla\psi)^2-J(\psi)(\nabla \theta - q A )^2 - V(\psi) \right)
\end{equation}
We require that the coupling functions $G$ and $J$ and the potential $V$ be even functions of $\psi$, since we need to preserve our $U(1)$ symmetry. We will assume an expansion of the form
\be\label{smallphi}
V=-{6} + m^2\psi^2+\Op(\psi^4), \quad G=1+ g\psi^2+ \Op(\psi^4), \quad J=\psi^2+\Op(\psi^4)
\ee
 where we have set the $AdS_4$ radius to one. The coefficient of $\psi^2$ in $J$ is fixed by regularity at $\Psi = 0$. This is all that we will need to determine the behavior near the critical point in section 3. We will specify the potential and coupling functions more fully later when they are needed to construct the ordered phase away from the critical point.

The mass $m$ of
the field around the symmetric point $\psi = 0$ determines
the conformal dimension of the dual operator $\mathcal{O}$ in the 
$CFT_{2+1}$ 
\begin{equation}
\label{massdim}
\Delta_\pm = 3/2 \pm \sqrt{ 9/4 + m^2 } 
\end{equation}
This also controls the asymptotic behavior of the scalar field
\begin{equation}
\label{falloff}
\psi(r) =  {\alpha\over r^{\Delta_-}} + {\beta\over r^{\Delta_+}} +\ldots \quad
{\rm as} \quad r \rightarrow \infty
\end{equation}
with the metric asymptotically approaching
\be
ds^2 = r^2 (-dt^2 + dx_idx^i) + {dr^2\over r^2}
\ee

In order for us to be able to add a double trace operator as in (\ref{eq:dblUV})
in a controlled fashion (without destroying the $AdS_{4}$ asymptotics) we require that the mass
 be in the range:
\begin{equation}
 - 9/4 < m^2  < -5/4
\end{equation}
In this range, both sets of modes are normalizable and one has a choice of boundary conditions  for quantizing the bulk theory. \emph{Standard} quantization corresponds to setting $\alpha = 0$, and  $\mathcal{O}$ has dimension $\Delta _+$. \emph{Alternative} quantization corresponds to $\beta = 0$, and $\mathcal{O}$ has dimension $\Delta_-$ \cite{Klebanov:1999tb}. We will refer to these two theories as $AdS_{4}^{\rm (std.)}$ and $AdS_{4}^{\rm (alt.)}$. We want to start with alternative quantization, so the dimension of $\mathcal{O}$ is
\be
 1/2 < \Delta_- < 3/2 \label{eq:relevantrange}
\ee
and add a double trace deformation. In this range, adding the double trace operator  (\ref{eq:dblUV}) amounts to studying
the gravitational theory in asymptotically $AdS_{4}$ space with new boundary 
conditions for the scalar \cite{Witten:2001ua,Berkooz:2002ug} 
\begin{equation}
\label{bc}
\beta = \kUV \alpha
\end{equation}
Up to an overall normalization, $\alpha = \langle \mathcal{O} \rangle$. Note that $\kUV$ has dimension $\Delta_+-\Delta_-=3-2\Delta_-$ and is thus a relevant coupling
in the range of interest (\ref{eq:relevantrange}). Hence adding the perturbation will induce an RG flow from the
original CFT to a new theory in the IR 
which can be understood by sending $\kUV \rightarrow \infty$. Dividing (\ref{bc}) by $\kUV$ as we take this limit 
we see that we arrive at $\alpha=0$, the theory with {standard} boundary conditions.
If we had started with $\kUV = 0$ we would stay at the unstable
fixed point in {alternative} quantization. 
Many details of this flow have been studied (see, e.g., \cite{Gubser:2002vv}).

\subsection{Symmetry breaking from double trace deformations}

We will first study a simple model of spontaneous symmetry breaking. We will work with zero chemical potential, and hence can set the bulk Maxwell field to zero. This is natural in a theory which preserves charge conjugation. 
Adding a double trace term (\ref{eq:dblUV}) with negative coefficient is expected to destabilize the vacuum. This is easily seen as follows.
Consider the two-point function for the operator $\Op$ in the vacuum (at zero chemical potential, temperature, and double-trace coefficient.) The retarded Green's function is (up to overall normalization)
\be
G^R_\pm(p)=p^{2\Delta_\pm-3} \, \qquad p^2 = - ( \omega + i \epsilon)^2 + \vec{p}^2
\ee 
where the sign $+(-)$ indicates the Green's function for the standard (alternative) quantized theory. Also $\omega$ and $\vec{p}$ are the energy
and momentum respectively. If we start with the alternative quantized theory and add a double-trace term of the form (\ref{eq:dblUV}), one finds\footnote{This can be calculated either in the field theory at large $N$ by summing up a geometric series of diagrams, or on the gravity side with proper treatment of boundary conditions.}

\be
G^R(p)=\frac{1}{\frac{1}{G_-^R(p)}+\kUV}+\Op(1/N^2).\ee
For $\kUV>0$, this just introduces a new massive pole with a width at 
$p_{pole}^2=(-\kUV)^{1/(3-2\Delta_-)}$. However, for $\kUV<0$, we have a tachyonic instability, with a pole at real positive $p_{pole}^2$. All of this is directly analogous to a massless free scalar field getting a massive deformation of either sign. In \cite{Ishibashi:2004wx} an exponentially growing tachyonic mode was explicitly found in the bulk precisely when $\kUV$ had the wrong sign\footnote{Strictly speaking, \cite{Ishibashi:2004wx} studied the theory on a sphere, where coupling to background curvature induces a positive double trace term for scalars. In that case, $\kappa$ needed to be sufficiently negative to find the instability. Since we are studying the theory on Minkowski space, the critical point is simply when $\kappa$ changes sign.}. It was widely believed that theories with  $\kUV<0$ would not have a stable vacuum, but in \cite{Faulkner:2010fh} it was proven that for many scalar gravity theories, there was a stable ground state with nonzero $\alpha = \langle\Op\rangle$.

The stability of the dual gravitational system depends on the global existence of a superpotential $P_c(\psi)$, and the zero temperature broken symmetry ground state is given entirely by $P_c(\psi)$. The details can be found in \cite{Faulkner:2010fh}. The key result is the behavior of the off-shell potential $\mathcal{V}(\alpha)$. 
It turns out that\footnote{The overall factor of 2 (which was not present in  \cite{Faulkner:2010fh}) arises since we do not have a $1/2$ in front of our action (\ref{action}).}
\be\label{defeffpot}
\V=2(\Delta_+ - \Delta_-) (W+W_0),
\ee
$W(\alpha)$ is given by our boundary conditions $\beta = W'(\alpha)$, and for our double trace deformation is simply $W(\alpha)= \kUV\alpha^2/2$.
  $W_0(\alpha)$ is found from a scaling limit of smooth horizonless static solitons in global $AdS$, again see \cite{Faulkner:2010fh} for details. At $T=\mu=0$, scale invariance implies 
\be\label{defsc}
 W_0(\alpha)=\frac{s_c\Delta_-}{3} |\alpha|^{3/\Delta_-}.\ee
The coefficient $s_c$ depends on the full bulk potential $V(\psi)$, and is generally positive. 
If $s_c$ is negative the theory is somewhat sick since $W_0$ is unbounded, and the alternative quantized theory is unstable, as it has states with arbitrarily negative energy. We do not consider
this case any further.
The full off shell potential is thus
\be\label{formofV}
\V(\alpha)= (\Delta_+ - \Delta_-)\left (\kUV\alpha^2+\frac{2s_c\Delta_-}{3} |\alpha|^{3/\Delta_-}\right).
\ee
\begin{figure}[h!]
\begin{center}
\includegraphics[scale=.7]{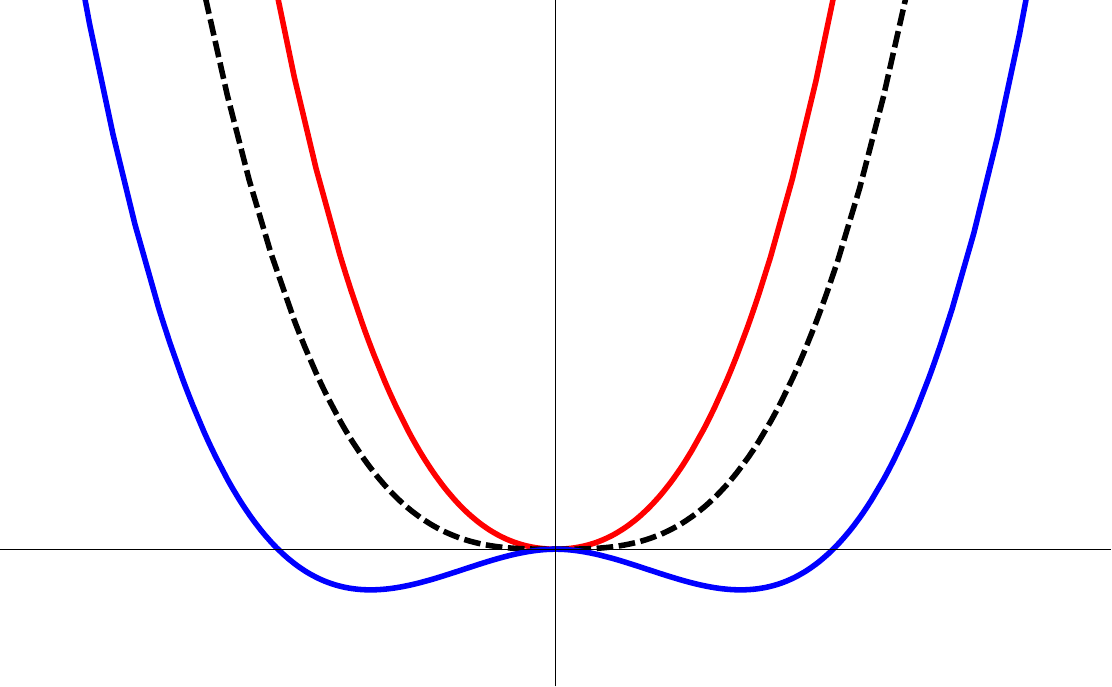}
\end{center}
\vspace{-.5cm}
\caption{
\label{fig:SSBpotential} The off-shell potential as we tune $\kUV$. The black dashed curve is the fine-tuned theory with $\kUV=0$, the blue curve is $\kUV<0$, and the red curve is $\kUV>0$. This is a strongly coupled version of the standard Landau-Ginzburg symmetry breaking mechanism.}
\end{figure}
For our range of interest (\ref{eq:relevantrange}) the second term dominates at large $\alpha$, and we have a classic example of spontaneous symmetry breaking with a saddle-shaped potential for negative $\kUV$ (see Fig. \ref{fig:SSBpotential}). The ground state which minimizes (\ref{formofV}) has
\be\label{neutralscaling}
\alpha = \langle \Op\rangle = \left(-{\kUV\over s_c}\right)^{\Delta_-/(3-2\Delta_-)}, \quad \V_{min}=-{1\over 3}\left(\frac{3-2 \Delta_-}{s_c^{\Delta_-/(3-2\Delta_-)}}\right)^2 (-\kUV)^{3/(3-2\Delta_-)}.
\ee
As mentioned above, the gravitational description of this ground state is uniquely determined by the superpotential $P_c(\psi)$.

Putting the theory at finite temperature can lift this instability. The detailed calculation is in appendix \ref{SSBTc}, but the result is that as we heat the system up to 
\be
\label{k1}
T_c =k_1 (-\kUV)^{1/(3-2\Delta_-)}, \quad {\rm with} \ k_1 = \frac{3}{4\pi} \left(-\frac{\Gamma((\Delta_+-\Delta_-)/3)\Gamma(\Delta_-/3)^2}{\Gamma((\Delta_- -\Delta_+)/3) \Gamma(\Delta_+/3)^2} \right)^{1/(3-2\Delta_-)}
\ee
the system returns to the symmetry preserving state. Note that everything scales as a power of $\kUV$, since this is the only scale in the problem.
Another way of studying this system at $T > 0$ is to construct the finite temperature generalization of $\V$. This can be obtained from the family of hairy black holes in the bulk, as described in \cite{Hertog:2005hu}.

\subsection{A novel holographic superconductor}

We saw above that adding a renormalizable double trace coupling can break a $U(1)$ symmetry at low temperature. This provides a new mechanism for  constructing holographic superconductors. Unlike the previous approach, which required a nonzero charge density to generate a low temperature condensate, we can now work at zero net charge. In this case, the Maxwell field remains strictly zero, even when the charged scalar hair is present in the bulk.

\begin{figure}[h!]
\begin{center}
\includegraphics[scale=.7]{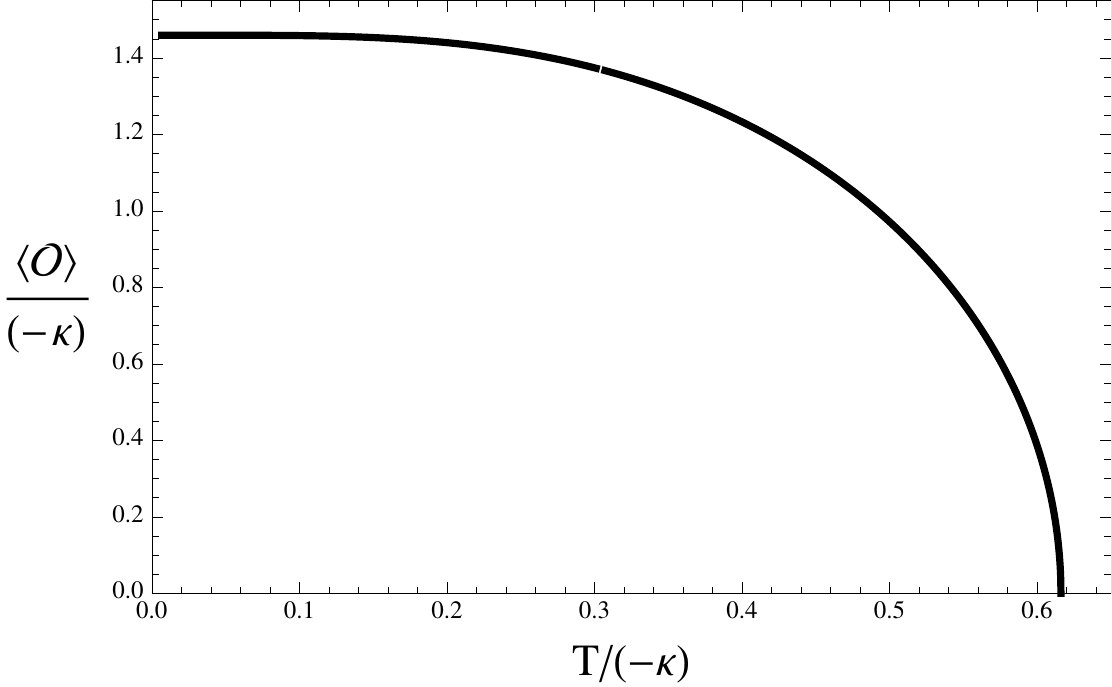}
\includegraphics[scale=.7]{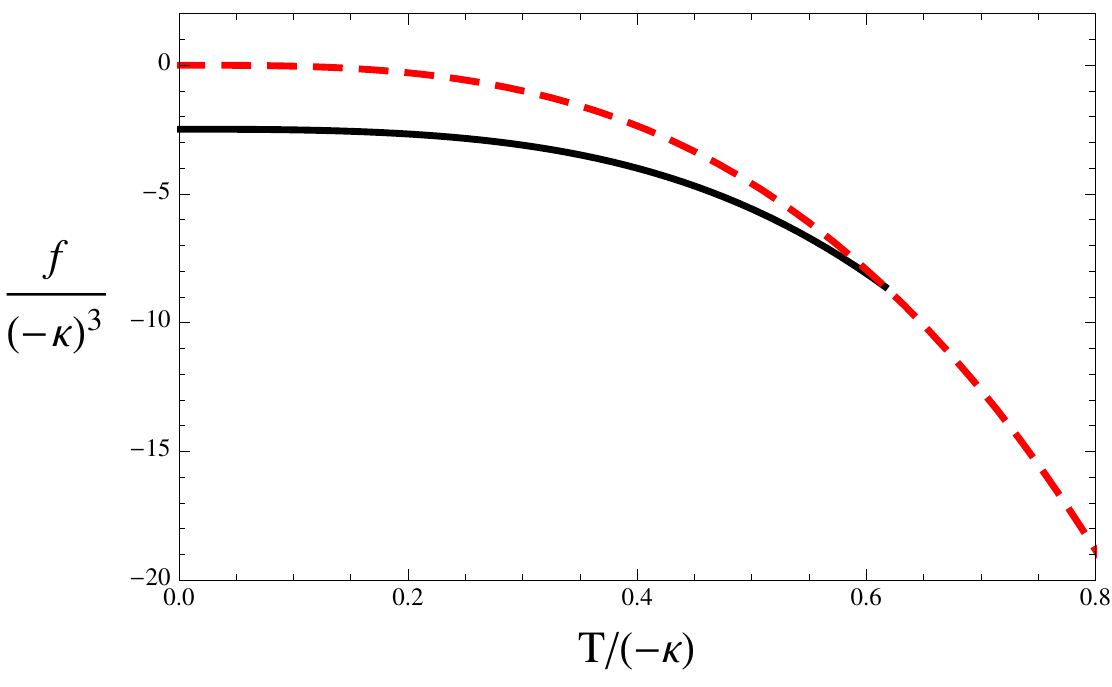}
\end{center}
\vspace{-.5cm}
\caption{
\label{fig:SSBexample} The order parameter $\alpha = \langle \Op\rangle$ and free energy density $f$ across the second order phase transition down to zero temperature. In the figure on the right, the dashed red line is the free energy of the normal phase (Schwarzschild AdS solution) and the black line is the free energy of the condensed phase.  We used a case with $\Delta_-=1$,  and bulk potential $V(\psi)=\sinh^2(\psi/\sqrt{2})(\cosh(\sqrt{2}\psi)-5)=-6-2\psi^2+\Op(\psi^4)$  \cite{Gauntlett:2009dn, Gubser:2009gp}. In \cite{Faulkner:2010fh} it was found that $s_c=0.56$ for this potential. From (\ref{k1}), we have $T_c/(-\kUV) \approx 0.62.$ 
 }
\end{figure}

We computed the critical temperature above. To study the system away from $T_c$, we need to specify the full nonlinear bulk potential. Working in units of $\kUV$ (which is analogous to working in units of $\mu$ or $\rho$ in cases of finite density), we find that the order parameter behaves just as it does in the case where we find an instability by lowering $T/\mu$ in the standard holographic superconductor setup. We can also calculate the difference between the free energy of the hairy black hole and the normal black hole, which is simply AdS-Schwarzschild, and find generically that below the critical temperature the hairy black hole is always preferred. An example which comes from a consistent string theory truncation is shown in Fig. \ref{fig:SSBexample}.

As usual, to compute the conductivity, one starts by perturbing the Maxwell field and metric. As shown in  Appendix \ref{appendix:EOM}, the conductivity can be simply related to the reflection coefficient in a one dimensional Schrodinger problem:
\be
-b''(z)+V_{Sch}(z)b(z)=\omega^2b,\label{schrodinger}
\ee
where $\delta A_x = a_x e^{-i\omega t}$ and $b=\sqrt{G(\psi)}a_x$. The Schrodinger potential is bounded (and given explicitly in (\ref{schrodpot})). $z$ is a new radial coordinate that vanishes at infinity and goes to minus infinity at the horizon. To obtain the required ingoing wave boundary condition at the horizon, we assume $b = e^{-i\omega z}+\R e^{i\omega z}$ near $z=0$ so that  $b=\mathcal{T} e^{-i\omega z}$ near the horizon. The conductivity is simply given by
\be
\sigma =\frac{1-\R}{1+\R}.
\ee
Since the potential is bounded, this will produce the usual behavior of the optical  conductivity. At low temperature there will typically be a gap at frequencies below the height of the potential\footnote{Since the Schrodinger potential is no longer positive definite, one can sometimes get peaks at low frequency in the conductivity \cite{Cadoni:2009xm}.}, and at higher frequencies the conductivity will approach its normal state value. There will be a delta function at $\omega = 0$ in the condensed phase, which can be seen from a pole in the imaginary part of the conductivity.  

The key difference from the holographic superconductors at nonzero charge density, is that there is no delta function in ${\rm Re}[\sigma]$ at $\omega = 0$ in the normal phase. This awkward feature of the previous construction arose since a state with net charge can be boosted, yielding a nonzero current with zero applied electric field. This implies infinite DC conductivity. Since we can now start in a state with zero charge, we no longer have this problem. Mathematically,  the delta function arose since the Schrodinger potential was nonzero in the normal phase due to a contribution from the background electric field. Here, the Schrodinger potential vanishes in the normal phase. The background is just the Schwarzschild AdS black hole, and $\sigma = 1$ with no delta-function contributions.

\subsection{Non-zero density and stability conditions}

In addition to providing another way to construct holographic superconductors, the addition of a double trace perturbation provides a new knob for adjusting the critical temperature of traditional holographic superconductors. As discussed in the introduction,  adding a term with $\kUV > 0$ makes it harder to condense the operator $\Op$ and lowers the critical temperature. We will see below that in some cases, $T_c \rightarrow 0$ as $\kUV $ approaches a finite value, $\kappa_c$. This is a new quantum critical point which will be studied in detail in the next section. 

With a nonzero charge density,  the normal phase is described by the Reissner-N\"ordstrom-AdS (RN-AdS) black hole. The critical temperature is determined by looking for a static normalizable mode of the scalar field in this background \cite{Denef:2009tp}. This marks the onset of the instability to form scalar hair. This problem only requires the leading terms in the functions $V(\psi), \ G(\psi), \ J(\psi)$ given in (\ref{smallphi}). The addition of the double trace term changes the critical temperature since it changes the boundary condition on the normalizable mode.

For reference, the RN AdS black hole is described by the metric and
gauge potential,
\bea
ds^2=  -f dt^2 + r^2 d \vec{x}^2 +\frac{dr^2}{f}\,, \quad \phi &=& \mu -\frac{\rho}{r}\,, \quad f = r^2 -\frac{m_0}{2 r }+\frac{\rho^2}{4r^2} 
\label{rnbh} \\
m_0 = \frac{2 \rho^3}{\mu^3} + \frac{ \rho \mu}{2} \,,
\qquad&&  T = \frac{\mu}{4\pi}  \left( \frac{3 \rho}{\mu^2} - \frac{\mu^2}{4\rho}  \right)
\eea
The horizon is located at $r_0 = \rho/\mu$, where $\rho$ and $\mu$
are the charge density and chemical potential respectively.

On this background the linear fluctuations are given by,
\begin{equation}\label{psieq}
\left( r^2 f \psi' \right)'  =   \left(    m^2 r^2 +  \vec{p}^2  -
\frac{ g \rho^2}{2 r^2} - \frac{ r^2 ( \omega + q \phi)^2 }{f} \right) \psi
\end{equation} 
where we have included momentum dependence $\vec{p}$ and
frequency dependence $\omega$ and $\Psi =\psi e^{ - i \omega t + i \vec{x} \cdot \vec{p} }$.
We will need these in Section 3
but for now they can be set to zero. Before going onto the case
of double trace boundary conditions, we first recall some
known results on stability conditions.  One natural question to ask is what is the condition
for the absence of an instability at any $T$. 
In other words, when is the normal state stable at zero temperature?
A necessary condition was
identified in \cite{Denef:2009tp}: the extremal RN black hole
in the near horizon limit becomes $AdS_2 \times R^2$, so
if the effective mass of $\psi$ derived
from (\ref{psieq}) is below the $AdS_2$ BF bound,
the RN BH will be unstable below some critical temperature.
So one condition for stability is demanding $m_{\rm eff}^2 > -1/4 $ where
\be
\label{stab1}
 m_{\rm eff}^2=\frac{m^2}{6} - \frac{q^2}{3} -g 
\ee
This is equivalent to 
demanding the conformal dimension
of the operator dual to $\psi$ in the $AdS_2$ region, $\delta_{\pm}$, 
are real, where
\be\label{defdelta}
\delta_{\pm} = \frac{1}{2} \pm \sqrt{\frac{1}{4} + m_{\rm eff}^2} \,.
\ee

This condition is too weak however, and we would like to refine it.
As we will see below, the stability condition $m_{\rm eff}^2 > -1/4 $ is  sufficient
for standard boundary conditions for the scalar $\alpha =0$ ($\kUV = \infty$).
However since alternative boundary conditions $\beta=0$ ($\kUV = 0$)  are weaker,
the scalar field can still be unstable to forming
hair despite the BF bound in $AdS_2$ being satisfied. This is because for any effective mass, there are always unstable modes in $AdS_2$. It is just that they are usually thrown out by the boundary conditions. With alternative boundary conditions in the asymptotic $AdS_4$ region, some of these unstable modes are allowed.

As further indication of the fact that alternative boundary conditions
are more unstable, it was noticed in \cite{Denef:2009tp} that $T_c$ diverges as one approaches the unitarity bound $\Delta_- = 1/2$. As shown in Fig. 3, this divergence actually takes the form 
\be
T_c \sim {\mu q\over (\Delta_- - 1/2)^{1/2}} 
\ee
Interestingly, for neutral scalar fields  there is no divergence and $T_c$ approaches a finite limit as $\Delta_- \rightarrow 1/2$.

\begin{figure}[h!]
\includegraphics[scale=1.]{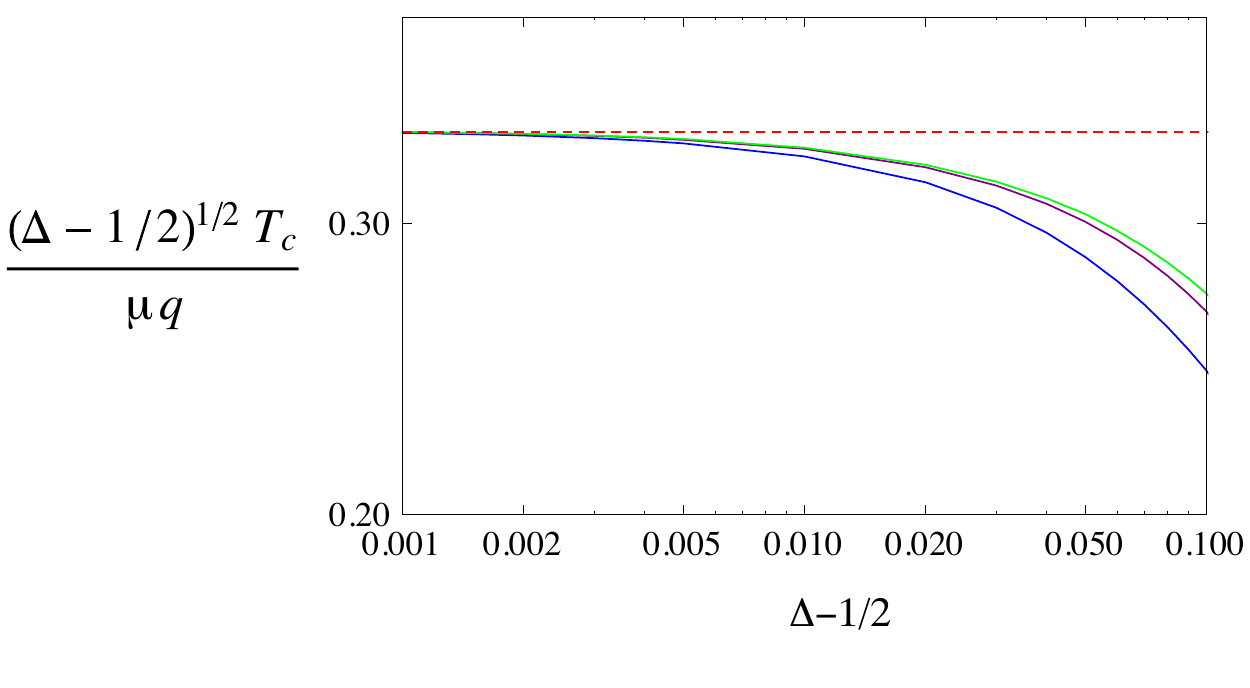}
\caption{\label{fig2}  The critical temperature as a function
of $\Delta_-$ for $q=.1, .25, .5$}
\end{figure}

\begin{figure}[h!]\begin{center}
\includegraphics[scale=1.]{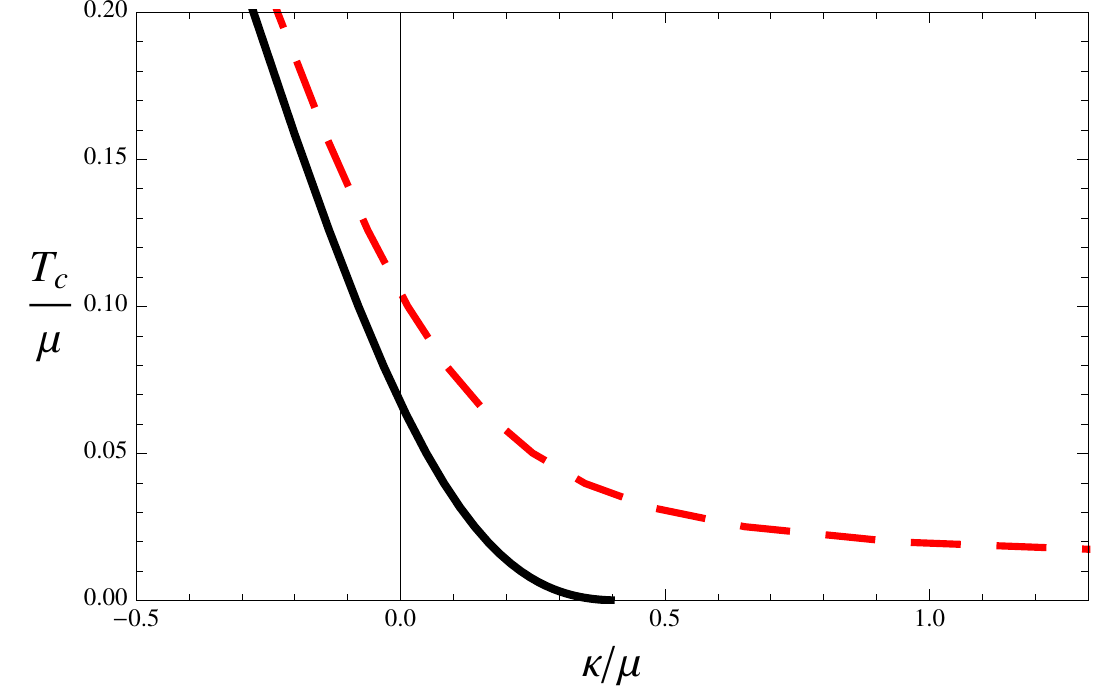}
\caption{\label{Tc_vs_k} The critical temperature, in units of chemical potential, as a function of the UV double trace coupling $\kUV$
for fixed $\Delta_-  = 1$ and $q=1/2$. The top curve has $g=0.2$ and has nonzero critical temperature for all $\kUV$. The lower curve has $g = -0.2$ and ends at a quantum critical point. }\end{center}\end{figure}

We  now want to include the effect of nonzero $\kUV$. 
Working at fixed $\mu$ the relevant scale invariant quantity that
we will vary is $\kUV/\mu^{\Delta_+ -\Delta_-}$ (as well
as $T/\mu$).
It turns out to be easy to study $T_c$ as a function of $\kUV$
simply by changing the definition of ``normalizable''.\footnote{Amusingly this
is a much simpler problem than the usual shooting problem.  
Rather than adjusting $T$ to find a static normalizable mode we can
simply fix $T=T_c$, shoot to the boundary and read off $\kUV(T_c)$.} 
Increasing $\kUV$ always decreases $T_c$. If the mass and charge of the scalar field is such that extreme RN AdS is unstable with standard boundary conditions, then $T_c$ remains nonzero for all $\kUV$. However, if extreme RN AdS is stable with standard boundary conditions, then $T_c$ must vanish at a finite value $\kUV = \kappa_c$. Both cases are illustrated in Fig. \ref{Tc_vs_k}.
Since $T_c$ can be arbitrarily large at $\kUV = 0$ and vanish at $\kappa_c$, we see that $\kUV$ is a very sensitive knob to adjust the critical temperature of the superconductor.
The point $\kUV= \kappa_c$ is the quantum critical point that we
will study in the next section. Notice that as $\kappa$ becomes large and negative in Fig. \ref{Tc_vs_k}, $\mu$ becomes less important, and both curves approach the scaling (\ref{k1}). 

\sect{Quantum critical point}

We now turn to a more precise discussion of the point $\kappa = \kappa_c$ where
the critical temperature $T_c / \mu \rightarrow 0$. 
Below we will outline the various requirements on
bulk parameters to achieve this critical point. Note in particular we need
not fine tune these bulk parameters. The coupling that we do tune $\kappa/\mu^{\Delta_+ -
\Delta_- }$
is a well defined boundary theory coupling.   

\subsection{The flow of double trace couplings and the 2 point function}

It is useful to think of the extremal  RN black hole
as representing a flow from a $CFT_{2+1}$ in the UV to a $CFT_{0+1}$ in the IR. 
This flow is induced in the UV by turning on a source  ($\mu$) for the charge density
operator $J^t$. The IR CFT can be seen by taking a scaling limit of (\ref{rnbh}) 
towards $r\rightarrow r_0$
at extremality.\footnote{This limit can be taken more carefully, keeping a finite but small $T$. 
See Appendix C. The scaling limit was discussed in \cite{Faulkner:2009wj} and the
discussion here follows that paper closely.} Rescale coordinates as:
\begin{equation}
\label{rescale}
 \hat{r} = \epsilon (r-r_\star)  \qquad \hat{t} = \epsilon t \qquad
 \hat{x} = r_\star x
 \end{equation}
where $r_\star$ is the location of the horizon at extremality $r_\star \equiv r_0|_{T=0} = \rho_\star/\mu$ with $\rho_\star \equiv  \rho|_{T=0} =  \mu^2/\sqrt{12}$.
Formally we can scale towards $r \rightarrow r_\star$ by expanding in $\epsilon$ then
setting $\epsilon=1$. This yields
\be
\label{ads2}
ds^2_0= \left( -  6 \hat{r}^2 d\hat{t}^2 +  \frac{d \hat{r}^2}{6 \hat{r}^2} \right)+  \left( d\hat{x}^2 + d\hat{y}^2 \right) , \quad ~ A_0 =  \sqrt{12} \hat{r} d\hat{t}
\ee
This is the classic $AdS_2 \times R^2$ geometry  found in the IR of
many extremal black hole solutions.
Notice in particular that the scale $\mu$ has dropped out. Since this
geometry is supposed to be dual to a scale invariant theory, this had to be the case. The only
knowledge that this theory has of the scale $\mu$ is encoded
in higher order irrelevant terms which we have dropped. For example keeping 
the next order terms in the expansion in $\epsilon$ one finds:
\begin{equation}
\label{irrel}
ds^2 = ds^2_0 + \delta_h  ds^2_1 + \ldots , \quad ~ A = A_0 + \delta_h  A_1+ \ldots
\end{equation}
where $ds^2_1$ and $A_1$ have energy scaling dimension $1$ under the $AdS_2$ 
scaling. The chemical potential now appears through $\delta_h = 1/\mu$.  
Note that since $\delta_h$ has  dimensions
of $-1$ this represents an \emph{irrelevant} coupling, which when turned
on induces a flow in the UV to $AdS_4$. It is useful to 
think of $\delta_h$ as opening up the $R^2$ directions of the metric.

We now return to linearized fluctuations of $\psi$ in the extremal RN 
background. Our goal will be to compute the two point 
function of the order parameter at small frequencies and momenta $\omega, p \ll \mu$. We proceed
heuristically, leaving details to Appendix C. Following \cite{Faulkner:2009wj} 
we do a matched asymptotic expansion where
we split the geometry into two regions. In both regions
we do a perturbative expansion in $\epsilon$ where
we redefine
\begin{equation}
\label{small}
\omega \rightarrow  \epsilon \omega \,,\quad \vec{p} \rightarrow \epsilon \vec{p}
\end{equation}
so as to access small frequencies and momenta.  In the \emph{inner}
region we rescale coordinates as in (\ref{rescale}). In the \emph{outer}
region we leave the coordinates unscaled. A systematic
expansion in both regions is defined in this way, matching
occurs in an intermediate region connecting the two. 
 
At zeroth order the outer region simply follows from setting $\omega = 0, T=0, \vec{p}=0$ 
in the full RN background. A general solution to the resulting equation
can be characterized by the behavior at the $AdS_4$ boundary and at the extremal
horizon,
\begin{equation}
\label{exp0}
 \psi(r \rightarrow \infty )  \rightarrow  \alpha_0 r^{-\Delta_-} + \beta_0 r^{-\Delta_+} \,, \qquad
\psi(r \rightarrow r_\star ) \rightarrow  \hat{\alpha}_0 (r-r_\star)^{-\delta_-} + \hat{\beta}_0 (r-r_\star)^{-\delta_+} 
\end{equation}
\begin{equation}
\label{matrix}
\begin{pmatrix}  \alpha_0 \\ \beta_0 \end{pmatrix} = L \begin{pmatrix} \hat{\alpha}_0 \\ \hat{\beta}_0 \end{pmatrix} \equiv \begin{pmatrix}
a^+ & a^- \\ b^+ & b^- \end{pmatrix} \begin{pmatrix} \hat{\alpha}_0 \\ \hat{\beta}_0 \end{pmatrix}
\end{equation}
where $a^\pm$ and $b^{\pm}$ are constants (in units of $\mu$) and 
can only be computed numerically.\footnote{ In \cite{Faulkner:2009wj} these same constants
were called $a_\pm^{(0)},b_\pm^{(0)} $.} 
If we impose linear boundary
conditions (\ref{bc}) on the allowed fluctuations, then this maps 
into the following condition near $r_\star$,
\begin{equation}
\frac{\hat{\beta}_0}{\hat{\alpha}_0} = \frac{(a^{+})^2 ( \kappa - \kappa_c)}{\det L - a^- a^+ (\kappa - \kappa_c ) } \quad \mathrm{where} \quad \kappa_c \equiv b^+/a^+  \quad
\mathrm{and} \quad \det L = \mu^2 \frac{ \delta_+ - \delta_-}{\Delta_+ - \Delta_-}
\end{equation}
We would like to make the identification of $\hat{\beta}_0/\hat{\alpha}_0$ above
with the value of a double trace coupling in the IR $AdS_2$ CFT,
$\kIR$. 
We will only be interested in $\kappa$ close to $\kappa_c$ such
that,
\begin{equation}
\kIR = \frac{(a^+)^2}{\det L} ( \kappa - \kappa_c) 
\end{equation}
It is then natural to identify $\kappa = \kappa_c$ as the critical
point that we observed in the previous section. One main
reason for this identification is the fact that for $\kappa < \kappa_c$
the double trace coupling in the IR is negative, and thus
analogous to the discussion in Section 2, there will
be a new state with lower free energy and scalar hair. 

We are now in a position to complete the computation
of the two point function of the order parameter. The retarded
Green's function follows from imposing incoming
boundary conditions at the extremal horizon in the inner region. Then
to zeroth order in the $\epsilon$ expansion 
one finds that $\hat{\beta}_0/\hat{\alpha}_0 = \Sigma_R(\omega)$
where $\Sigma_R$ is the retarded $AdS_2$ Green's function for
fluctuations on the background (\ref{ads2}).  The Green's function in the full
CFT can be computed using any of the usual prescription \cite{Son:2002sd,Iqbal:2008by}
generalized to include nonstandard boundary conditions. The
result is up to overall normalization,
\begin{equation}
\label{greenfull}
\chi_R (\omega, \vec p) = \frac{ \alpha}{- \beta+ \kappa \alpha} =  
\frac{ Z + \ldots }{  \kIR - \Sigma_{R}(\omega)+ \mathbb{X}(\omega,\vec{p}) + \ldots }
\qquad Z = (a^+)^2/\det L 
\end{equation}
where we have included higher order terms that can be
important in $\mathbb{X}$. These higher order
terms always come from perturbative corrections in the outer region and are thus
real. In contrast, $\Sigma_R$ is in general complex.
The elipses above represent even higher order
terms that we have dropped. We compute $\mathbb{X}$ in Appendix C.
The result can be written as,
\begin{equation}
\label{X}
\mathbb{X}(\omega, \vec{p}) =
c_p \vec{p}^2 - c_\omega \omega^2
 - c_T \kappa_c T + c_q q \left(- \omega + \frac{2 \pi}{\sqrt{3}} T q \right)
\end{equation}
where $c_i$ are constants in units of $\mu$. They have
the following positivity constraints depending on the value of $\delta_-$: 
$c_p >0, c_T > 0$ always, $c_\omega >0 $ for $\delta_- < -1/2$ and $c_q > 0 $ for $\delta_- < 0$.

The $AdS_2$ Green's function plays the role of the
self energy in (\ref{greenfull}) and is given by
\begin{equation}
\label{sigR0}
\Sigma_R(\omega, T=0) = h e^{i \phi}(- i  \omega)^{1 - 2\delta_-}
\end{equation}
where $h$ is a real positive number, and
$e^{i \phi}$ is a phase, the precise form of which does not matter.
We can generalize the above discussion
to finite but small $T$. At finite temperature $\Sigma_R$ takes the form
of a nontrivial scaling function,
\begin{equation}
\label{scaleform}
\Sigma_R(\omega,T) =  \left( \frac{2 \pi T}{3} \right)^{\delta_+ - \delta_-}
\frac{ \Gamma(\delta_+ - \delta_-)\Gamma(\delta_+ - i q/\sqrt{3}) \Gamma(\delta_+
+ iq/\sqrt{3} - i \omega/(2\pi T) )}{\Gamma(\delta_- - \delta_+ ) \Gamma(\delta_- - i q/\sqrt{3}) \Gamma(\delta_-
+ iq/\sqrt{3} - i \omega/(2\pi T) )}
\end{equation}
Importantly the $AdS_2$ Green's function always satisfies
the constraint,
\be
\label{spec}
\omega \rm{Im} \Sigma_R(\omega,T) > 0
\ee
which is necessarily true for any bosonic spectral density.

Generally speaking since the quantities computed
in the IR $AdS_2$ geometry depend simply
on two numbers $q$ and $\delta_-$ we will call
these quantities ``universal''. Since they are associated
with a CFT this language seems appropriate. Other
quantities that come from the outer region such
as $a^{\pm}, b^{\pm}$ and the $c_i$ are ``nonuniversal'', they
can be computed only numerically.

\subsection{Properties of the critical point}

Given the two point function (\ref{greenfull}) we can now
understand the physics close to the critical point. First
we study the phase boundary in the $(\kappa,T)$ plane where the order parameter
condenses, or equivalently, where the correlation
length diverges. 

We again look for
a static normalizable mode at $T=T_c$ which manifests
itself as a zero frequency pole in (\ref{greenfull}) . Since
the instability kicks in first for the homogenous mode we can
take $\vec p = 0$. There are two cases depending
on the IR conformal dimension $\delta_-$. For $0<\delta_-<1/2$
we can ignore $\mathbb{X}$ altogether, however for $\delta_- < 0$
the analytic correction $\propto T$ is larger than $\Sigma_R \propto T^{1-2\delta_-}$.
Thus we find for small $\kIR$,
\begin{eqnarray}\label{critemp1}
(0< \delta_- <1/2) \quad T_c &=& k_2 \left( - \kIR \right)^{1/(1 -2 \delta_-)}  \\
(\delta_- < 0 ) \quad T_c &=&  k_3 (- \kIR) \label{critemp2}
\end{eqnarray}
where,
\begin{eqnarray}
k_2 = \frac{3}{2\pi } \left(- \frac{ \Gamma(\delta_- - \delta_+)}{ \Gamma(\delta_+ - \delta_-)}
\left| \frac{ \Gamma(\delta_- - i q/\sqrt{3})}{ \Gamma(\delta_+
- iq/\sqrt{3} )}  \right|^2 \right)^{1/(1-2\delta_-)}  \quad
k_3 = \frac{1}{-c_T \kappa_c+ c_q q^2 2\pi/\sqrt{3}}
\end{eqnarray}
Note that while $k_2$ is a universal
number (one to be compared with (\ref{k1})) $k_3$ is nonuniversal,
depending on quantities $c_T$ and $c_q$  defined in the \emph{outer} region. 
Indeed it is not clear the sign of $k_3$ is fixed.
Although $c_q >0$ and $c_T >0$ for the range
of dimensions of interest,  $\kappa_c$ does not have a fixed sign. Generically it seems
that for a critical point occurring with $\delta_- < 0$ and
$q=0$ then $\kappa_c <0$
so that $k_3$ is fixed to be positive in such a case. However we do not know a proof
of the positivity of $k_3$.  

\begin{figure}[h!]\begin{center}
\includegraphics[scale=1]{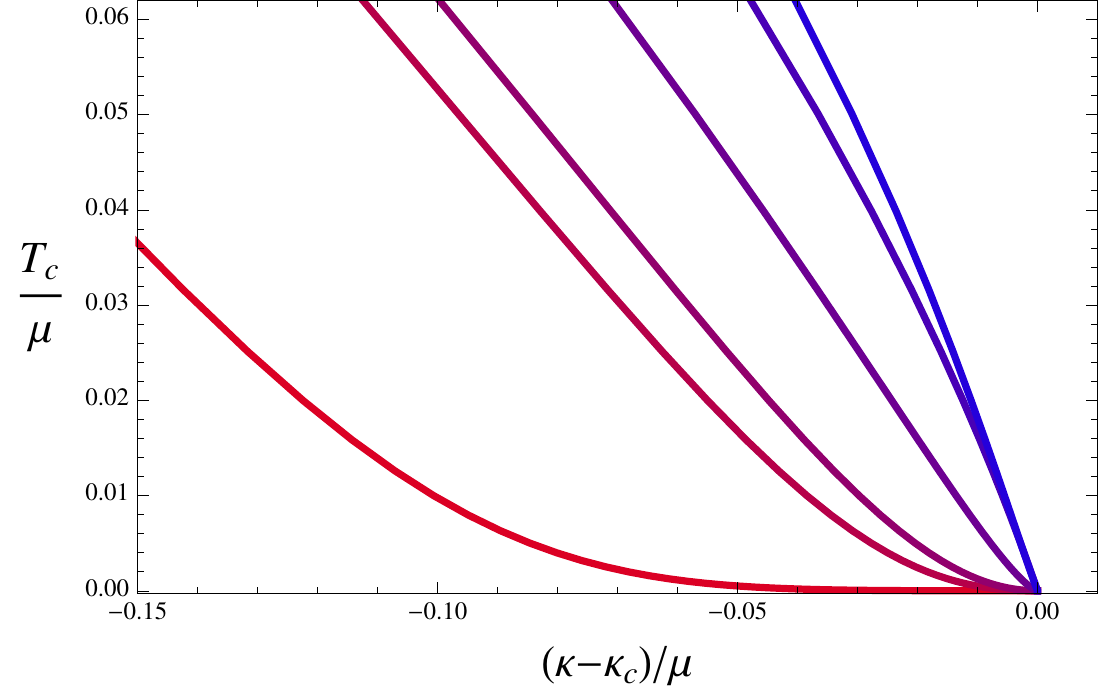}
\caption{\label{vary_Tc}
The critical temperature close to $\kappa_c$ for different values of $\delta_-$. These are for theories with $q=0$, $\Delta_-=1$, and from left to right:
$\delta_-=0.45,0.30,0.26,0.15,0,-0.15.$ Note that when $\delta_->0$ the critical temperature vanishes with a power law, but for $\delta_- \le 0$ it vanishes linearly.}
\end{center}
\end{figure}

To check the scaling relations (\ref{critemp1}) and  (\ref{critemp2}), we have numerically computed the critical temperature. In Fig.~\ref{vary_Tc} we plot $T_c$ as a function of $\kappa$ for various $\delta_-$. The results are perfectly consistent with our scaling relations.


Now we move away from the phase boundary and study
the disordered phase at zero temperature.
Here we can examine the structure of the retarded Green's function in the complex $\omega$ plane. 
We will be particularly interested in the dispersion of the mode
that becomes tachyonic on the ordered side. Again
depending on the value of $\delta_-$ the dispersion will differ, also
now there will be a difference if the order parameter is charged or not.
Examining (\ref{greenfull}) at $T=0$ for the \emph{charged} case the dispersion of the pole 
in the complex plane is,
\begin{eqnarray}
\label{critdisp}
(0< \delta_- < 1/2 ) \quad \omega_\star &=& i e^{-i \phi /(1 -2 \delta_-) } \left( \frac{c_p p^2 + \kIR }{h} \right)^{1/(1- 2 \delta_-)}  \\
\label{qneq}
( \delta_- < 0; q\neq 0) \quad \omega_\star &=& \omega_R(p) - \frac{\Sigma_R \left(\omega_R(p),T=0\right) }{ q c_q} \,,\qquad
\omega_R(p) =  \frac{ c_p p^2 + \kIR}{q c_q} 
\end{eqnarray}
where in the last case the width of the quasiparticle scales like $\rm{Im} \Sigma_R (\omega_R)
\propto |\omega_R|^{1-2\delta_-}$, which is smaller
than the energy $|\omega_R|$.  In this case we have a genuine quasiparticle
with a mass $ \propto \kIR$. In the first case (\ref{critdisp}) the width scales like
the energy and thus the pole does not represent a genuine quasiparticle.

For the \emph{neutral} case we again get the same
behavior as in (\ref{critdisp}) however now for the range of
conformal dimensions $-1/2  < \delta_- < 1/2$.
For the remaining neutral case,
\begin{equation}
\label{qzero}
( \delta_- < -1/2; q = 0) \quad \omega_\star^2 = \omega_R(p)^2 - \frac{\Sigma_R\left(\omega_R(p),T=0\right) }{ c_\omega}
\,, \qquad \omega_R^2(p) = \frac{  c_p p^2 + \kIR}{ c_\omega} 
\end{equation}
where there are now two quasiparticles at positive
and negative energies $\pm \omega_R$. The width of these quasiparticles
scales as $|\omega_R|^{-2\delta_-}$.

As expected, $\kIR$ is playing the role of a mass (or an energy gap.)
For all cases one can show via application of the constraint (\ref{spec})
that for $\kIR > 0$ the quasi particle pole always lies in the lower half
plane and there is no instability. There is 
a ``mass gap''\footnote{ This is not a genuine gap, since
there will always be gapless incoherent junk coming from the $AdS_2$
Green's function. However it does represent
a gap to the coherent part of the 2 point function, which
is represented by the dispersing pole. 
}, $E_g$, for all cases which is roughly given by the closest
approach of the pole to $\omega=0$.
For  the less universal cases
(\ref{qneq}) and (\ref{qzero}) one finds the expected
results, $E_g \propto \kIR$ and $\kIR^{1/2}$ for $q\neq0$
and $ q =0$ respectively. However for the  more interesting
``critical'' case the gap  scales as,
\begin{equation} 
E_g \sim  ( \kIR )^{\frac{1}{1-2 \delta_- }}
\end{equation}
The correlation length for all cases scales as $\xi \propto \kIR^{-1/2}$. 

When $\kIR < 0$, application of the constraint (\ref{spec}) shows
that the pole always lies in the upper half plane
for momenta $p < \sqrt{-\kIR/c_p}$, representing
an instability.

Exactly at the critical point $\kIR = 0$ we find a free gapless mode which disperses
in the complex plane as
\begin{equation}
\omega \sim |\vec{p}\, |^{z} \quad {\rm with} \quad z = \frac{2}{1 -2\delta_- }
 \end{equation}
for the ``critical'' case, and $z=2$ for $q\neq0$ and $\delta_- < 0$
and $z=1$ for $q = 0 $ and $\delta_- < - 1/2$. 
We thus conclude that the physics of the critical point
has a nontrivial dynamical critical exponent determined
by the dimension of an operator $\delta_-$ in the IR $AdS_2$ CFT.
This identification is consistent with the relationship
$E_g \propto \xi^{-z}$ in all cases. 
Interestingly $z$ has a lower bound in this model, with $z>2$
for the charged case and $z>1$ for the neutral case.

To summarize, in the
most interesting ``critical case'' where $0 < \delta_- < 1/2$
for $q\neq 0$ and $ -1/2 < \delta_- < 1/2$ for $q=0$
the two point function close to the critical point
has the universal scaling form,  
\be\label{univ}
\chi_R =  \frac{Z}{\kIR + c_p \vec{p}^2 + T^{2/z} g(\omega/T) }
\end{equation}
where $z =  \frac{2}{1-2 \delta_- } $ and $g(\omega/T)$
is a universal scaling function that follows from (\ref{scaleform}).
Note that since the correlation function
is analytic in $\vec{p}$ (the self energy is momentum independent) 
the critical point is `locally' quantum
critical \cite{si1}. This type of criticality goes beyond the usual Landau Ginzburg
paradigm often applied to quantum critical points, due
to the existence of the locally critical modes associated
with $AdS_2$.  These results are compatible with experimental
measurements of the spin susceptibility in a heavy fermion
compound $ CeCu_{6-x} Au_x $ at criticality \cite{exp}. In
order to compare the spin susceptibility to the two point function
of the triplet staggered order parameter (which is effectively $\chi_R$), one must shift
the momentum in (\ref{univ}) by the ordering vector associated to the anti-ferromagnetic
order $\vec{p} \rightarrow \vec{p} - \vec{K}$. Very similar results were also found theoretically
for the spin susceptibility in \cite{si1,si2}. 
The most important feature for comparing to experiments
was $\omega/T$ scaling of the susceptibility at the ordering vector
$\vec{p} = \vec{K}$ and a nontrivial exponent
$z \approx 2.7$ or $\delta_- \approx .13$.
Indeed we capture both features here,
although since our $\delta_-$ does not take a universal
value, it is hard to make a prediction for this exponent
without a real string embedding where $\delta_-$ will be fixed.

\subsection{Renormalization group interpretation}

We would like to now give an RG interpretation of
the above results and in so doing try to understand
what to expect of the zero temperature ordered phase when $\kIR < 0$.

We have already discussed a major aspect of the RG flow: the extreme
RN black hole represents a flow from one $2+1$ CFT in the UV
to a $0+1$ CFT in the IR. We would like to now understand
how this picture changes in the presence of the scalar $\psi$. 
For this purpose there is clearly a set of a 
distinct cases depending on the value of $\delta_-$ the conformal
dimension of the operator dual to $\psi$ in the IR CFT. 
Firstly for $\delta_-$ complex the IR CFT will never
be realized and there will be no critical point. We do 
not consider this case further here.
There are two remaining cases $0 < \delta_- <1/2$
and $\delta_- < 0$ which we turn to now.

\subsubsection*{ \underline{$ 0 < \delta_- < 1/2$}}

This range of dimensions is in the ``critical'' range identified
above where the 2 point function takes the more universal
form (\ref{univ}) for both neutral and charged cases. Here
we argue that this result is universally controlled by
the $AdS_2$ theory supplemented by double trace deformations.
The bulk field $\psi$ in the $AdS_2 \times R^2$ geometry is dual
to a  set of operators $\Psi_{\vec p}$ where $\vec{p}$ labels
the momentum in the $R^2$ direction, which one should
think of as a charge under the KK reduction down to $AdS_2$. $\Psi_{\vec p}$ can be viewed as the image of $\Op(t,\vec p)$ under RG flow.

The operator dimension of $\Psi_{\vec p}$ in the $0+1$
dimensional CFT is given
by $\delta_{\pm} + \mathcal{O} (p^2)$, where in
this range of dimensions we can take either value. The two point
function of $\Psi_{\vec p}$ is then,
\be
\Sigma_R^{\pm} \propto \omega^{2\delta_\pm -1} 
\ee
where the previously defined $AdS_2$ Green's function (\ref{sigR0}) is $\Sigma_R^+ = \Sigma_R$.
If we take the dimension
of $\Psi_{\vec p}$ to be  $\delta_-$ such that we are working in alternative quantization
then we can reproduce the result of (\ref{univ}) at $T=0$
simply by including the following deformation 
of the $0+1$ CFT \cite{Faulkner:2010tq}, 
\be
S_{0+1} \rightarrow S_{0+1} - \int dt \frac{d^2\vec{p} }{(2\pi)^2} 
(\kIR + c_p \vec{p}^2 )  \Psi_{\vec{p}}^\dagger \Psi_{\vec{p}}
\ee
Note for example when the deformation is zero, $\kIR + c_p \vec{p}^2 = 0$, the two point
function (\ref{univ}) scales as $\omega^{2\delta_- -1}$ consistent
with $\Psi_{\vec p}$ having dimension $\delta_-$. 

At zero momentum there is a single relevant double
trace coupling $\kIR$ at the critical point, that does
not explicitly break the symmetry. Just as in the
$AdS_4$ case discussed in Section 2, a positive
$\kIR$ will induce a flow to standard quantization
where the operators $\Psi_{\vec p}$ now have dimension $1-\delta_-= \delta_+$.
On the other hand, a negative $\kIR$ will induce an instability
which will lead to a symmetry
broken state. Since $c_p >0$ the zero momentum
mode will go unstable first, so in
 the ordered phase the homogenous mode $\Psi_{\vec{p} = 0}$
develops a vev. 

The ordered state can then be studied using gravity with the
bulk field $\psi$ turned on. For $\delta_- > 0$,
 $m^2_{\rm eff}$ defined in (\ref{stab1}) is
negative which means that for the disordered phase the scalar $\psi$
is sitting at a maximum of an effective potential; in the ordered phase
it will roll away from this maximum. The resulting geometry will 
tell us where the theory ends up in the deep IR. In fact there
are many possibilities which will depend on
the bulk functions $V(\psi), G(\psi)$ and $J(\psi)$. One 
possibility which we highlight in the next section is the theory
flows to a different $\wAdS_2 \times R^2$ geometry,
where the field $\psi$ sits at the minimum of a particular
effective potential to be discussed later. In Section 4 we will
outline some other possible examples of deep IR geometries. At this stage
to be appropriately noncommittal we call this gometry $X$. 

\begin{figure}[h!]
\begin{center}
\includegraphics[scale=.8]{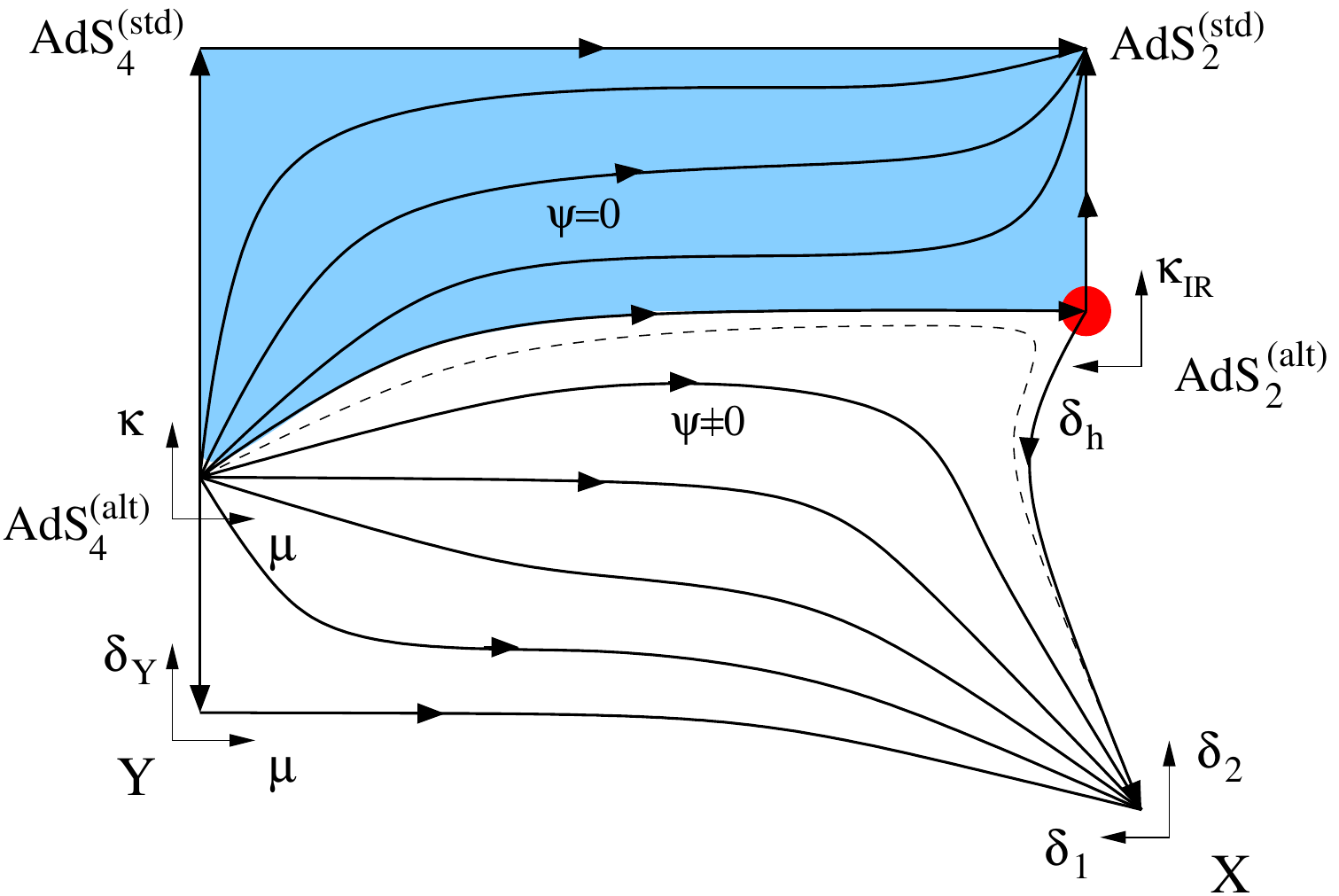}
\end{center}
\vspace{-.5cm}
\caption{
\label{fig4} Renormalization group flow interpretation
of the quantum critical point (denoted by the red dot)
for $0< \delta_- < 1/2$. This picture is heuristic, since
the couplings are only well defined close to each fixed point.  The blue
region is the disordered phase which we have studied
in this section. A distinction is made between
the theory in alternative (alt) and standard (std) quantizations.
The white region denotes the ordered phase
which we will study more carefully in the next section. 
The  theories denoted by $X$ and $Y$ depend on bulk
couplings. $Y$ is the end point of the flow induced
by turning on a negative double trace coupling in the $AdS_4$
theory and setting $\mu = 0$.  These
flows were considered in Section 2. $X$ is discussed
in the text above. 
 }
\end{figure}

The above discussion can be understood within
the context of the
 UV completion of $AdS_2$ (a fancy
name for the RN black hole) 
where we identify the coefficient of the double trace operator $\kappa$ as
the \emph{microscopic} control parameter which allows us to probe the critical point. 
The critical value $\kappa_c$ is thus not special
from the perspective of the UV theory.
The blue region of Fig.~\ref{fig4} is a pictorial description of 
the RG flows represented by the (extremal) RN black hole phase
of the theory. In the large-$N$ limit, this flow is trivial, only
effecting the fluctuations of fields in the bulk through boundary
conditions. 
When $\kappa < \kappa_c$, the double trace coupling
runs negative in the IR theory, and the instability ensues. 
The full $AdS_4$ theory then allows us to view the end
point of this instability - the geometry in the extreme IR flows
to a new attractive fixed point $X$.  These flows are represented
by the white region in Fig.~\ref{fig4}.

Finally we make an argument as to how the order parameter behaves
in the condensed phase. The argument\footnote{A similar
argument appeared in \cite{Jensen:2010ga} in the context of a BKT type transition,
and related arguments appear in \cite{D'Hoker:2010ij}.} is based
on an RG analysis close to the critical fixed point
and relies on the picture given in Fig.~\ref{fig4}. We will
confirm the result with a numerical calculation in the next section.
Consider shooting from the theory $X$ in the extreme IR
up to $AdS_4$. As suggested by Fig.~\ref{fig4} there
will be two tuning parameters, of which only an appropriate
scale invariant ratio of the two will matter. As we tune this ratio
we can shoot closer and closer to the $AdS_2$ critical 
theory. The dashed line in Fig.~\ref{fig4} is an example
of such a flow. In the $AdS_2$ region, the metric and Maxwell field are given by (\ref{ads2}) and the scalar field takes the form
\begin{equation}
\label{core}
\psi(\hat r) \sim v \, \hat{r}^{- \delta_-} + w\, \hat{r}^{- \delta_+} 
\end{equation}
Here $v$ and $w$ are constrained by scale invariance:
\be
\label{defscir}
w = - s^{\rm IR}_c\, v^{\delta_+/\delta_-}
\ee
where $s^{\rm IR}_c$ is a number which can be computed numerically.
It is analogous to the $s_c$ in (\ref{defsc}).

We
can now match (\ref{core})  onto linear fluctuations in the extreme RN background which
we have considered above (what we called
the outer region above, see for example (\ref{exp0})). 
\begin{equation}
\label{match}
\hat{r} = \Lambda (r-r_\star) \,, \quad \hat{t} = t / \Lambda \, , \quad
\hat{\alpha}_0 = v \Lambda^{-\delta_-} \, , \quad
 \hat{\beta}_0 = w \Lambda^{-\delta_+}
\end{equation}
where we have allowed for an arbitrary rescaling $\Lambda$
which will drop out in the end.
Additionally we have for $\kappa$ close to $\kappa_c$,
\be
\alpha_0 = a^+ \hat{\alpha}_0 \,, \quad  \hat{\beta}_0 = \kIR \hat{\alpha}_0
\ee

Putting these results together we find,
\begin{equation}
\label{eq:order}
\langle \mathcal{O}\rangle = \alpha
\approx a^+ \left(-{\kIR\over s^{\rm IR}_c} \right)^{\frac{\delta_-}{1-2\delta_-}} \,, \end{equation}
This is precisely an IR analog of the scaling we derived in section 2 for a negative double trace perturbation (\ref{neutralscaling}). Note we have made an assumption that $s_c^{\rm IR}$ is positive.
This is a nontrivial assumption and will depend on the bulk
couplings, as for example was the case for the
analogous $AdS_4$ parameter $s_c$ \cite{Faulkner:2010fh}. 
However whereas in the $AdS_4$ case, a negative
$s_c$ meant a somewhat sick theory, we suspect
a negative $s^{\rm IR}_c$ will simply result in a first
order transition. Since we are working with
the assumption of a continuous transition we cannot say much about
the case with $s^{\rm IR}_c$ negative. 

More generally we can consider a black hole solution
with a non-zero temperature which has $\psi$ nonzero
and comes close to $AdS_2 \times R^2$. Again
this is a nontrivial shooting problem.
Now there is a one parameter
family of solutions labeled by the temperature $\hat{T}$ and asymptotic to
(\ref{core}). Scale invariance imposes the following relationship
between $v$ and $w$,
\begin{equation}
w = - S \left( \frac{\hat{T}}{ v^{1/\delta_-} }\right)\, v^{\delta_+/\delta_-}
\end{equation}
where $S$ is a scaling function with $S(0) = s^{\rm IR}_c$. It can only be computed
numerically. Going through the same matching procedure
and rescaling $\hat{T} = \Lambda T$ one finds
the following scaling relation,
\be
- \kIR =   \left( \langle \mathcal{O} \rangle/ a^+ \right)^{\frac{ 1-2\delta_-}{\delta_-}}
S\left( \frac{T}{(\langle \mathcal{O}\rangle/ a^+)^{1/\delta_- }} \right)
\ee

This is of course consistent with the dimension
of the IR operator which gets a vev 
\be
\left<\Psi_{\vec{p}} \right> = \hat{\alpha}_0 \delta^2(\vec p)
\ee
being $\delta_-$ in alternative quantization
and $\kIR$ having dimensions $1-2\delta_-$.
In the next section we will construct these flows on the condensed
side.  We will
confirm amongst other things the result (\ref{eq:order}).

Actually this story is incomplete, for $0 < \delta_- < 1/4$ 
the $AdS_2$ CFT has higher order multi-trace operators
that are relevant (first $\hat{\alpha}^4$ then $\hat{\alpha}^6$ etc.). In this
situation the scaling arguments above fail, since
the critical point is now multi-critical. Away from the multi-critical
point, on the continuous side we now expect a mean
field\footnote{We thank
Kristan Jensen for drawing our attention to this possibility.}
 relationship $\langle \mathcal{O}\rangle \sim ( - \kIR)^{1/2}$. The argument for this goes as follows. Nonlinear corrections $\mathcal{O}(\psi^3)$
to the linear equation for $\psi$ in the outer region
produce additional terms in the matching (\ref{match}).
Most importantly 
\begin{equation}
\label{nonlin}
\hat{\beta} = \hat{\beta}_0 + \hat{\beta}_{NL} 
+ \ldots \approx \kIR \hat{\alpha}_0 + u_{\rm IR} \hat{\alpha}_0^3 = w \Lambda^{-\delta_+}  \,,\quad
\hat{\alpha} = \hat{\alpha}_0 + \hat{\alpha}_{NL} 
+ \ldots \ \approx \hat{\alpha}_0 = v \Lambda^{-\delta_-}
\end{equation}
where $u_{\rm IR}$ can be computed along the lines
of Appendix C.  It is obvious that we should interpret
$u_{\rm IR}$ as the RG flow of the quadruple trace operator
from $AdS_4$ to $AdS_2$ analogous to $\kappa \rightarrow \kIR$.
Thus at zero temperature we find,
\be
 \kIR \hat{\alpha}_0 + u_{\rm IR} \hat{\alpha}_0^3 = 
 - s_c^{\rm IR} (\hat{\alpha}_0)^{(1-\delta_-)/\delta_-}
\ee 
For $\delta_- > 1/4$, the $u_{\rm IR}$ term is not
important for small $\hat{\alpha}_0$ and we reproduce
the result (\ref{eq:order}). However for $\delta_- < 1/4$
the non-analytic term in $\hat{\alpha}_0$ is less
important and we get the mean field answer assuming
that $u_{\rm IR} > 0 $,
\be\label{quartic}
\left<O\right>  = a^+\left( - \kIR/u_{\rm IR} \right)^{1/2}
\ee
for $u_{\rm IR} < 0$ we get a first order transition that we cannot
say much about.
These nonlinear corrections to the linear $\psi$ equation come from
potential terms, as well as from back-reaction on gravity. They
are rather complicated, however they are most likely
computable in the probe approximation introduced in \cite{Iqbal:2010eh} for 
the neutral case. We leave their explicit computation to future work.

\subsubsection*{ \underline{$  \delta_- < 0$}}

In this case, $m_{eff}^2$ defined in (\ref{stab1}) is positive, so the IR $AdS_2$ with $\psi = 0$ is stable. However, one can still have phase transitions which turn on $\psi$ at larger radius. In this case, we will see the critical exponents are not governed by the $0+1$ CFT but take mean field values.

 We can reproduce
the more general 2 point function (\ref{greenfull}) for $\mathbb{X}$ given 
in (\ref{X}) using the following semi-holographic action \cite{Faulkner:2010tq},
\bea
\label{wf}
S &=& S_{0+1} + \int d^3 x \left( |\partial_t \Phi|^2 +  q_\Phi \left( i \Phi^\dagger \partial_t \Phi  + {\rm h.c} \right)  - c^2 | \vec{\partial} \Phi |^2 - \kappa_\Phi | \Phi|^2 - \frac{1}{2} u_\Phi | \Phi|^4  \right) \\
&+& \eta \int dt \frac{ d^2 \vec{p} }{(2\pi)^2} \left( \Phi_{\vec p}^\dagger \Psi_{\vec p} + {\rm h.c}  \right)
\eea
where $\Phi$ is a boundary field which we have
coupled to the $AdS_2$ CFT operator $\Psi_{\vec p}$ and $\Phi_{\vec p}$
is the spatial fourier transform of $\Phi$. 
To reproduce  (\ref{greenfull}) 
the dimension of $\Psi$ must be $\delta_+ + \mathcal{O}(\vec{p}^{\,2} )$.
The two point function for $\Phi$ will then agree with $\chi_R$ with
the following identifications,
\be
|\eta|^2 = 1 / c_\omega \,, \quad \kappa_\Phi = \kIR / c_\omega
\,, \quad q_\Phi = q c_q / c_\omega \, \quad c^2 = c_p / c_\omega\,, \quad
\Phi \sim \mathcal{O}   \sqrt{c_\omega/Z}  
\ee
Note that $\mathcal{O}$ will also have an overlap with $\Psi$ but
this will lead to a subdominant correction to the two point function. 
We have also included a nonlinear interaction term $u_\Phi$
in (\ref{wf}) which should be generated in the flow from $AdS_4$ to $AdS_2$. 
We will assume that $u_\Phi >0$, but this need not be the case.

In order to construct the ordered phase we first assume
that we can treat $\eta$ perturbatively. We will
show this is a consistent assumption. So for now we
set $\eta=0$ and work
in the mean field approximation for $\Phi$. We do
this because we are working in the classical gravity
approximation. (The whole action
above should be multiplied by $1/G_N$, and for
small $G_N$ mean field applies.) Then for $\kappa_\Phi$ negative,
$\Phi$ develops a vev:
\be
\left< \Phi \right> = \sqrt{ - \kappa_\Phi/u_\Phi }
\ee
Turning on $\eta$, this will now act as a source
for the homogenous mode of the operator $\Psi_{\vec p}$,

\be
S = S_{0+1} + \eta \sqrt{ - \frac{ \kappa_\Phi}{u_\Phi } } 
\int dt \frac{ d^2 \vec{p} }{(2\pi)^2} \delta^{2}(\vec{p})  \left( \Psi^\dagger_{\vec p}  + {\rm h.c.} \right) 
\ee
Since the dimension of $\Psi$ is irrelevant ($\delta_+ > 1$)
this source will scale away in the IR. Thus we do not
expect the vev of $\Phi$ to back-react on the $AdS_2$ CFT
which is now a stable fixed point. This is consistent
with the fact that for $\delta_- < 0$, 
the effective mass square $m^2_{\rm eff} > 0$,
and the bulk field $\psi$ sits at a minimum in 
the disordered phase and has nowhere to go in the
ordered phase. Thus the ordering is controlled by
the boundary field $\Phi$. 

\begin{figure}[h!]
\begin{center}
\includegraphics[scale=.8]{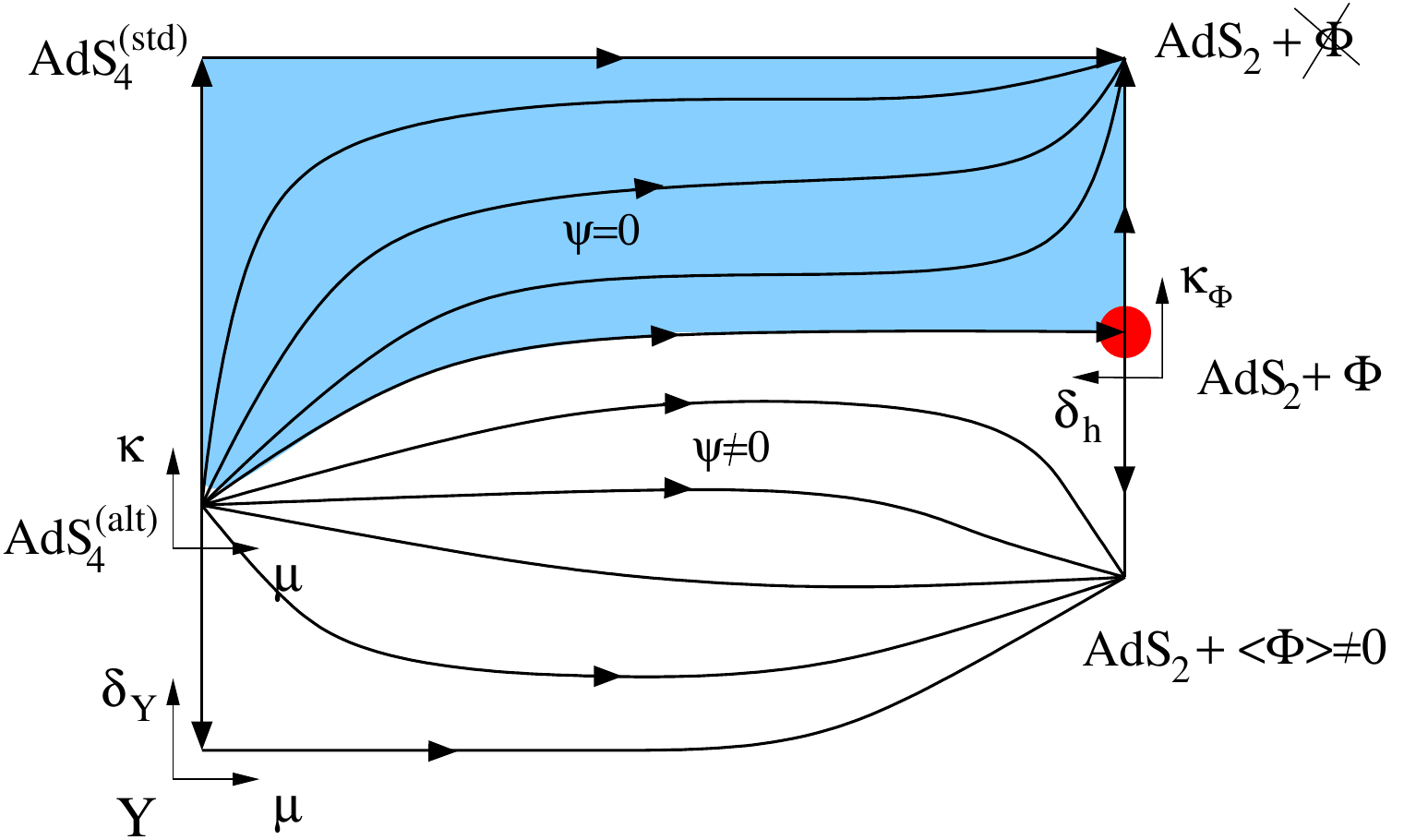}
\end{center}
\vspace{-.5cm}
\caption{ RG interpretation for $\delta_- < 0$. In this
case  $m_{\rm eff}^2 >0$, so
from the gravity perspective $AdS_2$ is stable and
the three fixed points  on the right side are all the same.
The condensation is controlled by an $AdS_2$ boundary
field $\Phi$. The three cases are distinguished by whether $\Phi$ has a nonzero vev (bottom), is part of the IR dynamics but does not condense (middle), or is massive so it drops out of the IR theory (top).
\label{RG2}  }
\end{figure}

It is now clear how to construct the ordered phase from gravity,
we simply shoot from $AdS_2$ with a non-zero
source term for the irrelevant operator $\Psi$. 
(One also needs to turn on the irrelevant coupling
$\delta_h$ discussed around (\ref{irrel}).)  This situation
is depicted in Fig.~\ref{RG2}. 
Again we can match this onto
the perturbations of the extreme RN black hole
with 
\be 
\hat{\alpha} \approx \hat{\alpha}_0 = \eta \sqrt{ - \kappa_\Phi/  u_\Phi}
\,, \qquad \hat{\beta} \approx \hat{\beta}_0 + u_{\rm IR} \hat{\alpha}_0^3 = 0
\ee
where we have included an important nonlinear correction as in (\ref{nonlin}).  
The condition $\hat{\beta} = 0$ (which is the requirement
that $\psi$ not blow up in the IR) and $\hat{\beta}_0 = \kIR \hat{\alpha}_0$
then allow us to match $u_\Phi$,
\be
u_\Phi = u_{\rm IR} / c_\omega^2 
\ee
The vev then follows from the value of the source term $\hat{\alpha}_0$,
\be
\label{mf5}
\langle \mathcal{O}\rangle = \alpha = (a^+/c_\omega) ( - \kIR/u_\Phi)^{1/2} 
 = a^+ ( -\kIR/u_{\rm IR} )^{1/2}
\ee
which is same as (\ref{quartic}).

\subsection{Parametric dependence on bulk couplings}

\begin{figure}[h!]
\begin{center}
\includegraphics[scale=.95]{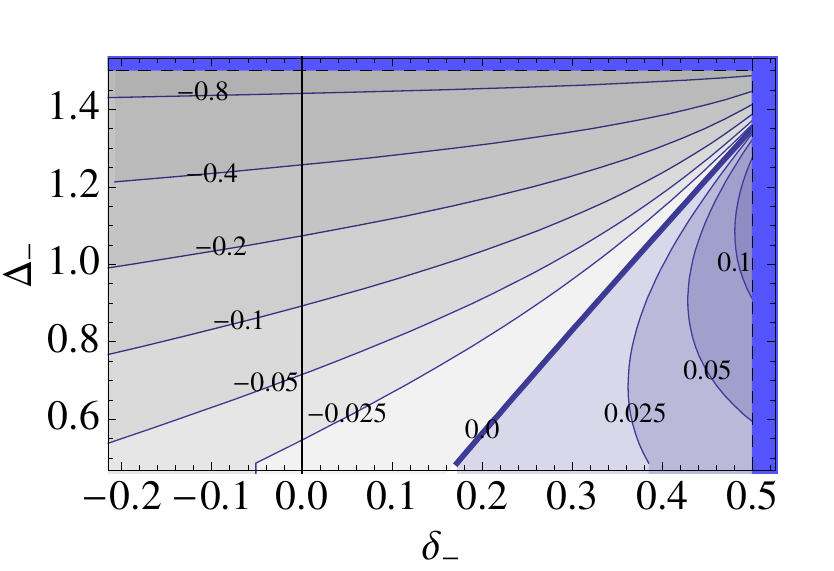}
\includegraphics[scale=.95]{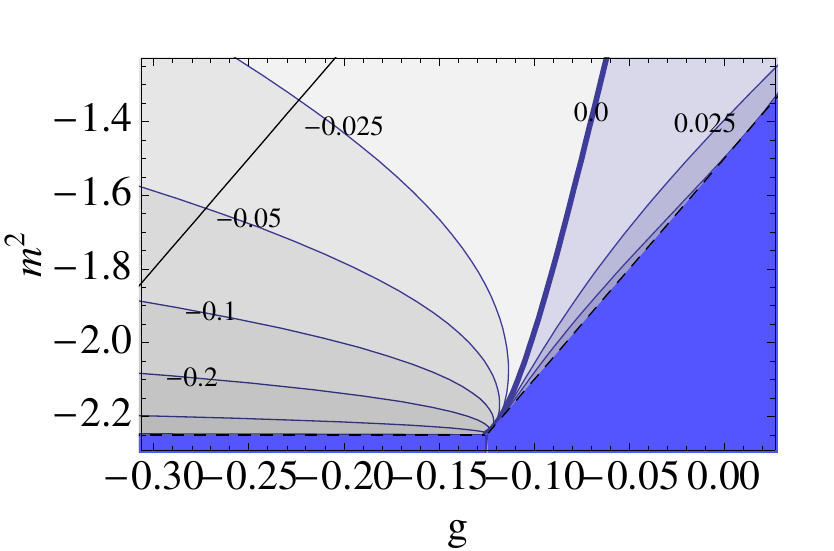}
\end{center}
\vspace{-.5cm}
\caption{\label{fig:ddmg}
Contour plots of $\kappa_c$ for $q=0$. The lines $\delta_- =0$ and $\kappa_c = 0$ are shown. The solid (blue)
region represents the excluded BF bound in $AdS_2$ and $AdS_4$.
Note that positive $\kappa_c$ tends to occur
close to this region since then the theory is more unstable.
For theories with $\kappa_c$ negative, introducing a chemical
potential fails to destabilize the theory with alternative boundary
conditions. So these theories are more stable.
}
\end{figure}

\begin{figure}[h!]
\begin{center}
\includegraphics[scale=.95]{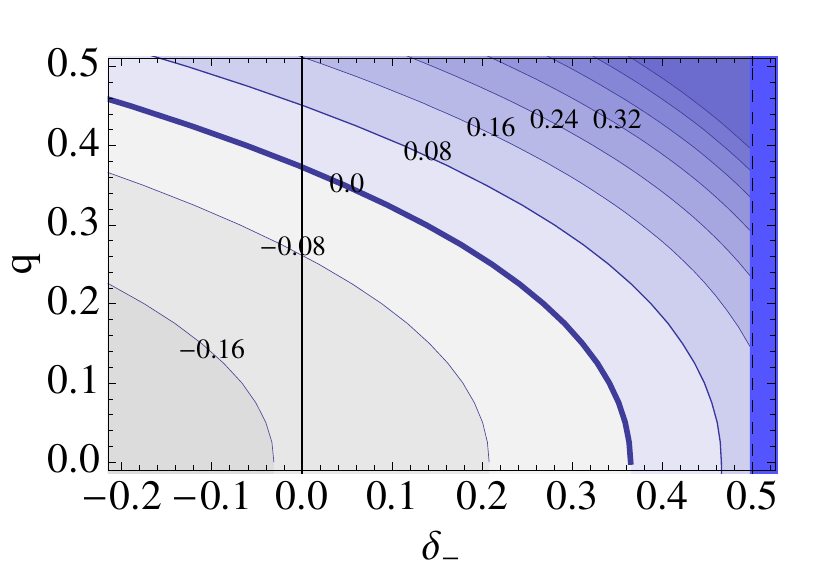}
\includegraphics[scale=.95]{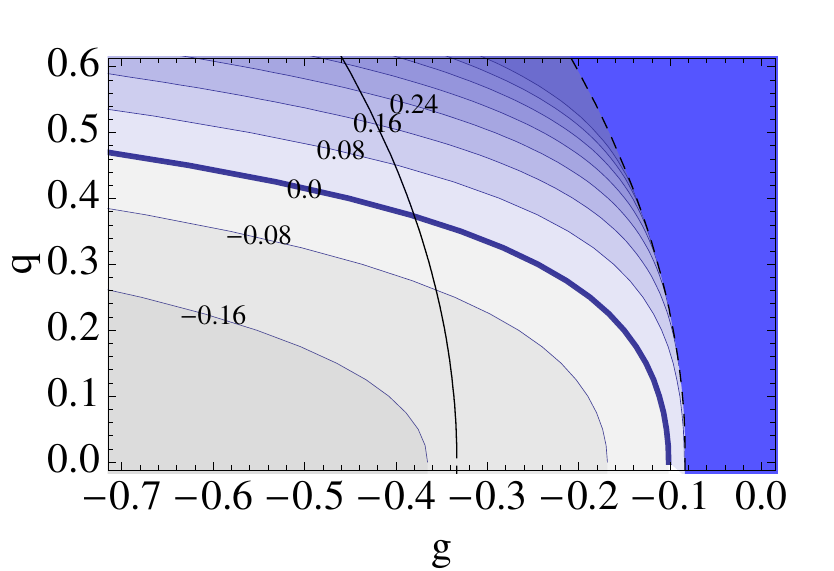}
\end{center}
\vspace{-.5cm}
\caption{\label{fig:qdqb}
Contour plots of $\kappa_c$ for $\Delta_-=1$. The lines $\delta_- =0$ and $\kappa_c = 0$ are shown. 
Clearly increasing $q$ tends to destabilize the theory
since $\kappa_c$ increases. Decreasing
the bulk coupling $g$ also tends to increase stability.
}
\end{figure}

In this section we compile some numerical results
relating to the critical point. The universal features
discussed above depend on two parameters $q$ and $\delta_-$.
However the location of the critical point itself $\kappa_c$
and other nonuniversal constants appearing in the dynamic
susceptability depend on three bulk parameters, $m^2, g , q$. 
In Fig.~\ref{fig:ddmg} and Fig.~\ref{fig:qdqb} we plot
$\kappa_c$ through two different slices of this three
parameter space, one with $q=0$ and the other with $\Delta=1$.

\sect{Constructing the ordered phase}

We have left many of the details of constructing the ordered  phase
to this section. We will consider the full back-reaction of
the condensing field $\psi$ on the metric 
so the problem is highly nonlinear and we must proceed
numerically. We will focus on the zero temperature
case where  the basic problem will be  
to find the appropriate IR solution and to integrate outwards from there. 
Perturbing around the IR solution one finds various shooting parameters that 
take the form of irrelevant couplings. These couplings generate flows
in the UV to an asymptotically $AdS_4$ solution, representing
the UV fixed point of the theory. From here we can read
off various data such as the vev of the order parameter,
the double trace couplings $\kappa$ and the free
energy.

As one tunes the irrelevant couplings in the IR we
find that one can shoot closer  
and closer to the critical point that we identified in the previous section.
As we will see this involves a flow whose geometry
is described, for a large chunk of radial proper distance, 
by the $AdS_2 \times R^2$ critical solution.

\subsection{Ansatz for the background}

We will use the following metric and field ansatz,
\begin{equation}
ds^2=- f(r) dt^2+\frac{dr^2}{f(r)}+h^2(r) d\vec{x}^2, \qquad
A = \phi(r) dt , \qquad \psi = \psi(r)\label{eq:ansatz}
\end{equation}
This metric differs from the one used in \cite{Hartnoll:2008kx} in which the radial coordinate was chosen to be $\sqrt {g_{xx}}$. The form (\ref{eq:ansatz}) is more convenient since it allows $AdS_2 \times R^2$ as a solution.
Also numerically it is more convenient to work with this metric
as we will demonstrate later.

The equations of motion which follow from the action (\ref{action})
are given in Appendix \ref{appendix:EOM} (\ref{modulusEOM}-\ref{einsteinEOM}). For our ansatz (\ref{eq:ansatz}) this reduces to the system of ODEs (\ref{eq:1}-\ref{eq:4}).
There are three important reparameterizations that leave the metric form invariant, 
thus allowing us to generate new solutions from old solutions. They are:
\begin{eqnarray} 
\label{eq:shift}
\textrm{Shift}: \quad r &\rightarrow &r +  a. \\
\label{eq:conf}
\textrm{Conformal rescaling}:  \quad r &\rightarrow& \Lambda r,   \,
(t,\vec{x}) \rightarrow (t,\vec{x})/\Lambda
,  \, f \rightarrow \Lambda^2  f,\nonumber
\\
\quad h& \rightarrow& \Lambda h, \, \phi \rightarrow \Lambda \phi.
 \\
 \label{eq:spatial}
\textrm{Spatial rescaling}: \quad \vec{x} &\rightarrow & \vec{x}/ s , \, 
h \rightarrow s h.
\end{eqnarray}
We will fix these reparameterizations by demanding certain asymptotic
boundary conditions, which we specify in the next subsection.

\subsection{Asymptotic $AdS_4$ data and the free energy}

The asymptotic UV fixed point will always be $AdS_4$.
For $\Delta_+ < 3$ there are no irrelevant (single trace) operators within our truncation 
so $AdS_4$ is always an attractive fixed point in the UV. A complete
expansion about $AdS_4$ can be systematically derived as discussed in 
  Appendix \ref{ap:FullAdS4}. 
The leading terms include 
\begin{eqnarray}
\nonumber
\psi &=& \alpha r^{- \Delta_- } + \beta r^{-\Delta_+} + \ldots  ,   \qquad 
\phi = \mu  - \rho  r^{-1} + \ldots \\
h &=& r \left( 1 + 0\, r^{-1} +  \ldots \right) ,  \qquad 
f h^{-2} =  \left( 1  -  (m_0/2)  r^{-3} + \ldots \right)
\label{ads4exp}
\end{eqnarray}
Additional terms in these expansions will be needed for some of the manipulations
which follow, however due to their cumbersome form we leave
these corrections to the appendix. 

A general solution is parameterized by 5 constants $\alpha, \beta, \mu, \rho, m_0$.
We have used the shift (\ref{eq:shift})  to fix the sub-leading $r^{-1}$ terms in the metric to zero, and used the spatial rescaling in  (\ref{eq:spatial})  to fix the normalization of $h$ and thus
the spatial components of the boundary metric. The remaining conformal symmetry (\ref{eq:conf})  can be used to fix
one of these five constants. We will work with fixed
chemical potential $\mu$, and often set  $\mu=1$.
 Numerical results of dimensionful quantities will be quoted in units of $\mu$. 

We will  need to compute the thermodynamic potential, which in our ensemble will be the grand potential $G$. We will
go through this in some detail by computing the Euclidean on-shell action.
While the results for general scalar boundary conditions are known 
using other methods the inclusion of a gauge field is new.
 
The grand potential is
 $G/T   = S_{E} + S _{ct}$ where $S_{E}$ is the Euclidean action and $S_{ct}$ 
 are boundary counter terms.   These counter terms are required
for a good variational problem, as well as to regulate divergences in $S_{E}$.
We want to keep the metric
fixed on the boundary (necessitating the Gibbons Hawking term), and the chemical potential fixed. 
For the scalar field $\psi$ we would like to require $\alpha, \beta$ to be constrained by 
 $\beta = W'(\alpha)$.  We will work initially with a fixed source $\beta$ in alternative quantization 
 where $W(\alpha) = \beta \alpha$, and generalize later.
The counter
terms in the case of fixed $\beta$, $S_{ct}^\beta$, and $S_E$ are given in Appendix \ref{ap:FullAdS4}, with the 
free energy density evaluating to:
 
 \begin{equation}
 g_{\beta}  =  - \frac{1}{2} m_0  +  \frac{2}{3} \alpha \beta \Delta_+ (\Delta_+ - \Delta_-)
 \label{free}
 \end{equation}
 where we have defined $g  = G/V$ and $V$ is the field theory volume.

We can also compute how the free energy varies as we move
in solution space parameterized by changes in the five integration constants $\delta \alpha, \delta \beta, \delta \mu, \delta \rho, \delta m_0$. 
Varying the on shell action we derive an analog of the first law of thermodynamics,
\begin{equation}
\label{eq:varf}
\delta g_{\beta} =  -  s \delta T - \rho  \delta \mu + 2 (\Delta_+ - \Delta_- ) \alpha \delta \beta 
\end{equation}
where $s$ is the entropy density. 
From (\ref{eq:varf}) it is clear that $g_\beta$ is stationary at fixed $T$, $\mu$, and $\beta$. 
To generalize the boundary conditions for $\psi$ we simply define 
\begin{equation}\label{defgW}
 g_W = g_\beta - 2 (\Delta_+ - \Delta_-) ( \alpha \beta - W) 
\end{equation}
who's variation is given by,
\begin{equation}
\delta g_W = - s \delta T - \rho \delta \mu -
2 (\Delta_+ - \Delta_- ) \delta \alpha(  \beta - W'(\alpha) )
\end{equation}
Notice that $g_W$ is stationary if $\beta = W'(\alpha)$ (and temperature and chemical potential
are fixed). The free energy $g_W$ is what we will be concerned with
in this section, the preferred state of the system must have lowest $g_W$.

To summarize, in order to have a well defined variational problem
and a finite on shell action, for boundary conditions determined by
an arbitrary function $W(\alpha)$ with fixed chemical potential, we require the following boundary
counter terms:
\begin{equation}
S_{ct}^W = S_{ct}^\beta - 2 (\Delta_+ - \Delta_-) \int d^3 x (\alpha \beta - W)
\end{equation}
where $S_{ct}^\beta$ is given in Appendix \ref{ap:FullAdS4}. 

At zero temperature we will not be minimizing the energy $\epsilon_W$
but rather the appropriate free
energy at fixed chemical potential $\epsilon_W - \rho \mu = g_W (T=0)$.  
To do this we present one final
manipulation which effectively fixes the constant $m_0$ 
in terms of the other integration constants. Consider
the free energy at fixed $\alpha$ (and $T=0$):
\begin{equation}
 g_\alpha =   g_\beta - 2 (\Delta_+ - \Delta_-) \alpha \beta
\quad \rightarrow \quad d g_\alpha = - \rho d \mu - 2 (\Delta_+ - \Delta_- ) \beta d\alpha
\label{ea}
\end{equation}
Using scale
invariance we can write $ g_\alpha = \mu^3 w( \alpha / \mu^{\Delta_-})$. 
Plugging this into (\ref{ea}) we find,
\begin{equation}
w'(a)  =  - 2 (\Delta_+ - \Delta_- ) b \,, \qquad
- \rho /\mu^2 =  3 w(a)  -  \Delta_- a w'(a)
\label{2eq}
\end{equation}
where we have defined $a = \alpha/ \mu^{\Delta_-}$ and $b = \beta/\mu^{\Delta_+}$.
The first equation above tells us that $g_\alpha$ can be found as an integral,
\begin{equation}
\label{abint}
  g_\alpha -  g_{RN}  =- 2 (\Delta_+ - \Delta_- ) \int_0^\alpha d\alpha' \beta(\alpha')
 \equiv   2(\Delta_+ - \Delta_-) W_0(\alpha)
\end{equation}
where the integral should be understood as occurring at fixed
$\mu$ (and $T=0$). When we turn off the scalar field altogether
$\alpha =0 ,\beta=0$ there is a unique solution: the
extremal RN black hole.  This fixes the integration
constant in (\ref{abint}).  Using (\ref{defgW}) and (\ref{ea}), the full
free energy density is
\begin{equation}
 g_W - g_{RN}   = 2 (\Delta_+ - \Delta_- ) \left[ W_0(\alpha) + W(\alpha) \right]
\end{equation}
Notice that the right hand side is identical to the off shell potential (\ref{defeffpot}).
The above discussion provides a thermodynamic derivation (generalized to nonzero chemical potential) of this result. 

The second equation in (\ref{2eq}) multiplied by $\mu^{3}$ becomes,
\begin{equation}
-\mu \rho =  3  g_\alpha (T=0)  + 2 \Delta_- ( \Delta_+ - \Delta_-)\alpha \beta
\end{equation}
Using (\ref{free}) and (\ref{ea}) we then find that $\mu \rho = (3/2) m_0$.
More generally with nonzero $T$, scale invariance imposes a thermodynamic
relationship which can be derived analogous to the above manipulations.
The result is,

\begin{equation}
m_0 = \frac{2}{3} \left( \mu \rho + T s \right)
\quad \rightarrow \quad
 g_W = -\frac{1}{3} (T s +\mu \rho)
+ 2(\Delta_+ - \Delta_-) \left( - \frac{1}{3} \Delta_- \alpha \beta + W(\alpha) \right)
\label{freeimp}
\end{equation}
The first equation was derived in \cite{Gubser:2009cg} using the Noether
charge associated with the conformal rescaling.
Practically speaking these last two equations are the most
useful for reading off the free energy from numerical data.
Since the $m_0$ coefficient is highly subleading
it is more accurate to use $\mu$, $T$ and $s$ to find $m_0$ using
the above equation.

\subsection{IR fixed point and shooting}

To begin with we would like to construct the zero temperature
ground state. One reason to do this is so we can see how the order
parameter $\left<\mathcal{O}\right>$ behaves as we tune
across the critical point.

To understand the ground state in the ordered phase we must find the final IR fixed point
geometry. This should not be confused with the $AdS_2$ critical point governing the behavior seen in section 3, but instead is the extreme infrared limit of the ordered phase we called $X$ in section 3. 
Depending on the specifics of one's model, there are many possibilities for infrared fixed points, and we will not attempt to classify all possibilities here. The case of unbounded $V(\psi)$ was studied in  \cite{Horowitz:2009ij}, which found various null singularities in the zero temperature limit. When the bulk potential $V(\psi)$ is regulated, in the
sense that it has a global minimum, the system is better behaved. The ground state
depends on whether the field is charged or not. For the
charged case, this problem was first considered in \cite{Gubser:2009cg}
where the possible IR solutions  were $AdS_4$ solutions
based about the new minimum or Lifshitz solutions.

In this section we will mostly focus on the neutral case $q=0$. Since
the field $\psi$ does not carry any charge, there
are no possible sources for the electric field lines. So
there are one of two possible outcomes, either the
horizon carries charge (the electric field is sourced on the horizon)
or the field $\psi$ runs to a point where the coupling
function $G(\psi)$ blows up\footnote{Another possibility, outlined recently in \cite{Balasubramanian:2010uw}, is that a curvature
singularity develops in the IR which sources the gauge field, one which is physically
sensible \cite{Gubser:2000nd}. Such behavior was seen for unbounded potentials at $q=0$ in \cite{Horowitz:2009ij}, though not thoroughly understood. This would be an interesting
state to consider since then the ordered phase could
be insulating (gapped to charged excitations.)}. Although this later
case would be an interesting possibility we will leave this problem to future work. For now
since we have a regulated potential and we are assuming
that $G(\psi)$ is smooth then this later case will not be realized.
In the former case since the horizon is charged, at zero temperature, there
must necessarily be an $AdS_2 \times R^2$ solution
in the IR, with constant scalar field
\cite{Iqbal:2010eh}. We will refer to this solution as $\wAdS_2$.

When $q=0$ the coupling function $J(\psi)$ does not
play a role. The equations of motion simplify greatly, reducing to
\bea
\phi' &=& \frac{\rho}{h^2G(\psi)},\nonumber\\
\psi''&+&\left(\frac{f'}{f}+\frac{2h'}{h} \right)\psi'-\frac{V'(\psi)}{2f}+\frac{\rho^2G'(\psi)}{4fh^4G(\psi)^2}=0,\nonumber\\
h''&+&h\psi'^2/2=0,\nonumber\\
\frac{h'^2}{h}&+&\frac{h'f'}{f}-\frac{h\psi'^2}{2}+\frac{hV(\psi)}{2f}+\frac{\rho^2}{4fh^3G(\psi)}=0,\nonumber\\
f''&+&\frac{2f'h'}{h}-\frac{\rho^2}{2h^4G(\psi)}+V(\psi)=0.\label{eq:neutraleom}
\eea
where the last equation is  a redundant dynamical equation. The first equation is easily integrated (with $\phi = 0$ at the horizon) after a solution to the other equations is obtained. A solution to (\ref{eq:neutraleom}) with constant scalar field and $AdS_2\times \mathbb{R}^2$ metric is easily found with the ansatz
\be
f=f_0 r^2, \quad h=h_0, \quad \psi=\psi_0.
\ee
The equations of motion reduce to
\be
f_0=-V(\psi_0), \quad \frac{\partial}{\partial \psi_0}\left(G(\psi_0)V(\psi_0) \right)=0, \quad \rho^2=-2 h_0^4G(\psi_0)V(\psi_0).
\ee
Note that the field $\psi$ sits at a minimum of an effective
potential 
\be
V_{\rm eff} (\psi) = V (\psi) G(\psi).
\ee
  This is our IR $\wAdS_2$ solution. To flow in the UV towards
$AdS_4$ we must add irrelevant perturbations. Since $\wAdS_2$ is stable, there are no relevant  perturbations
other than the one which leads to finite temperature. 

The method of studying deformations of fixed points is standard perturbation theory. Starting with a known fixed point $\Phi_0=(\psi_0,f_0,h_0)$, consider an expansion
\be
\Phi=\Phi_0+\delta \Phi_1+\delta^2\Phi_2+\ldots
\ee
 We call the perturbation relevant if the expansion is convergent at large radius and divergent at small radius, and irrelevant if it is divergent at large radius but convergent at small radius\footnote{The technical definition is that if the scaling dimension of the deformation is greater than the space-time dimension of the CFT, it is irrelevant and classically scales away in the infrared. If it is less than the field theory space-time dimension it is relevant, and becomes important in the infrared. }. Schematically, the equations of motion will be of the form
\be
D^2\Phi_1=0,~D^2\Phi_2=\Phi_1^2,\ldots
\ee
One can carry out this procedure to arbitrary order to increase the precision of one's numerics. Carrying this out for $\wAdS_2$ we find
\bea
\label{sm}
h&=&h_0+\epsilon( h_1r) +\Op(\epsilon^2),\nonumber\\
\psi&=&\psi_0+\epsilon\left(\tilde{\alpha}r^{-\wDelta_-} +\tilde{\beta}r^{-\wDelta_+} +c_0\,  h_1 r \right)+\Op(\epsilon^2), \nonumber\\
f&=&f_0 r^2+\epsilon \left(a + b r +c_1~h_1 r^3+c_2~\tilde{\alpha} r^{2-\wDelta_-}  +c_3~\tilde{\beta} r^{2-\wDelta_+}\right) + \Op(\epsilon^2),
\eea
where
\be
\wDelta_\pm = \frac{1}{2}\pm\sqrt{\frac{1}{4}+\frac{h_0^4 V_{eff}''(\psi_0)}{\rho^2}},
\ee
and the constants $c_0,~c_1,~c_2$ and $c_3$ are unenlightening constants fixed by the equations of motion. We follow the method of identifying relevant and irrelevant deformations in \cite{Gubser:2009cg} in terms of the $\wAdS_2$ scaling. Since $V_{eff}''(\psi_0)>0$, $\wDelta_+>1$, we see perturbing the scalar field is irrelevant, and so we will want to turn on the mode vanishing in the infrared, namely the $\walpha$ mode. 
Similarly changing $h$ is relevant, and we can turn on $h_1$. The deformations $a,b$ in $f$ correspond to turning on a finite temperature, which we do not wish to do. With this it is clear that $\wAdS_2$ is a totally attractive fixed point, and we therefore expect it to be the true ground state. To flow to $AdS_4$ at zero temperature we then only  want to turn on $ \walpha$ and $h_1$, since they vanish in the infrared. Thanks to the conformal symmetry we in fact only have one shooting parameter, $\walpha/h_1^{\wDelta_-}$, and we can use our scaling symmetries to work at fixed $\mu$. Integrating (\ref{eq:neutraleom}) to large radius with this small perturbation at small radius we will find an asymptotically $AdS_4$ solution. Note that we have only 
presented the shooting method (\ref{sm}) to leading order in perturbation theory. Often (when $\wDelta_+\gg 1$) it is necessary to work to higher order to have trustworthy numerical results.

\begin{figure}[h!]
\begin{center}
\includegraphics[scale=.95]{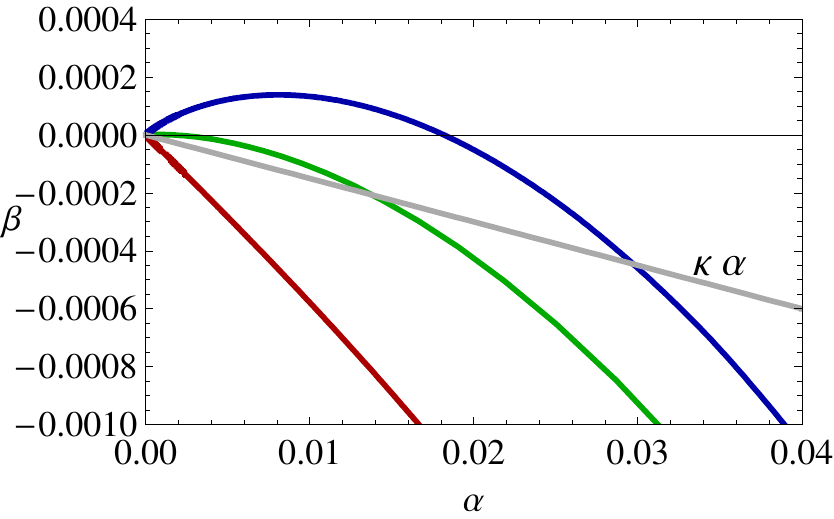}
\includegraphics[scale=.95]{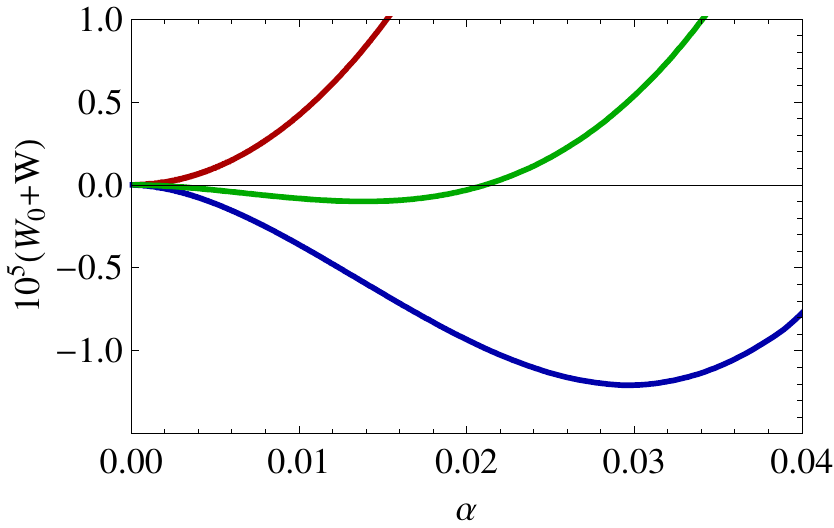}
\end{center}
\vspace{-.5cm}
\caption{\label{fig:ab}
The left figure shows $\beta(\alpha)$ curves computed
numerically as outlined above. The model used has
$V(\psi) = -6 + \Delta(\Delta-3) \psi^2 + \lambda \psi^4$ with $\Delta=1$, $\lambda=1$,
and $G(\psi) = \textrm{sech}[(-2g)^{1/2}\psi]$.  
The curves from top to bottom represent three
different choices of $- g = .085,.125,.140$.  These values
where chosen to give representative critical point conformal
dimensions. The gray line shows an example double trace
coupling $W' = \kappa \alpha$ for $\kappa= -.015$. 
The right figure
shows the effective potential $W_0 + W$ for
the same models when $\kappa=-.015$. Note that
the minima occur at the same points as the
intersection points in the left figure. Also note that
the difference in free energies in the broken phase is negative.
}
\end{figure}

\begin{figure}[h!]
\begin{center}
\includegraphics[scale=1]{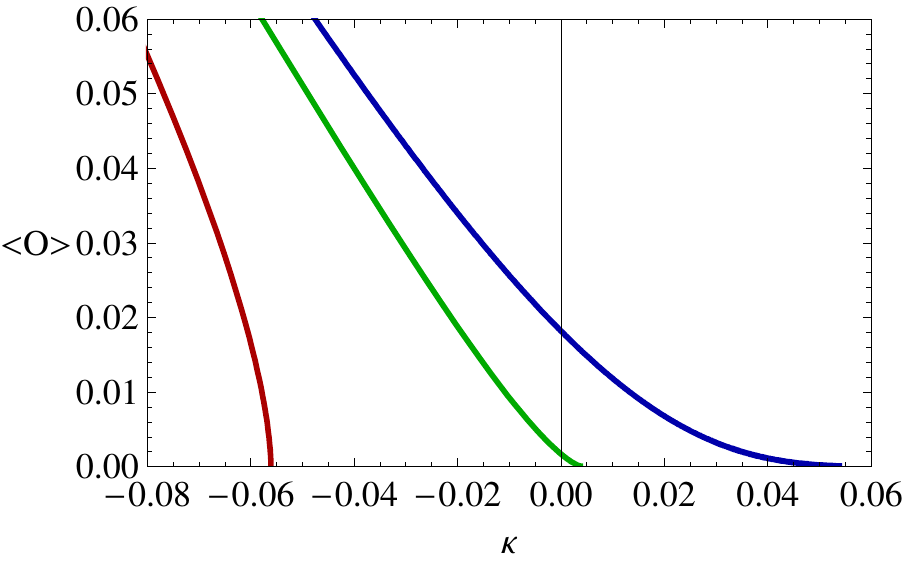}
\end{center}
\caption{\label{ak} This figure
represents a redrawing of the left panel of Fig~\ref{fig:ab}
using the relation $\kappa = \beta/\alpha$ and $\left< \mathcal{O} \right>=\alpha$.
The critical point $\kappa_c$ is where $\left< \mathcal{O} \right> \rightarrow 0$.
Interestingly these curves roughly
interpolate between mean field behavior and a transition
of infinite order (BKT transitions.)
}
\end{figure}

As we tune $\walpha/h_1^{\wDelta_-} $ we fill out a curve in the $\beta, \alpha$ plane.
We give some examples in Fig.~\ref{fig:ab}.  The basic strategy for reading off the possible states
for a given $W(\alpha)$ is to find the intersection
of the boundary condition curve $\beta = W'(\alpha)$ with the $\beta(\alpha)$  curves. 
One can then read off the vev of the operator $\left< \mathcal{O} \right>  = \alpha$
and the free energy using (\ref{freeimp}). For example, for a
 double trace deformation $W(\alpha) = (1/2) \kappa \alpha^2$,
we look along lines $\beta = \kappa \alpha$.
An example of the resulting $\left< \mathcal{O} \right>$ as a function of $\kappa$ is shown 
in Fig.~\ref{ak}.

\subsection{Confirming the scaling relations close to the critical point}

We now examine the above results close to the critical point. 
This point can be seen in Fig. \ref{ak} as the value of $\kappa = \kappa_c$
where $\left< \mathcal{O} \right>$ vanishes. As expected, this value of $\kappa_c$
agrees well with the calculation of where $T_c =0$ in section 2.  For $\kappa > \kappa_c$
there are no solutions with scalar hair, so the ground state
is simply the extremal RN black hole. For $\kappa < \kappa_c$ the free energy
of the ordered phase is smaller, so this is preferred.

\begin{figure}[h!]
\begin{center}
\includegraphics[scale=1.1]{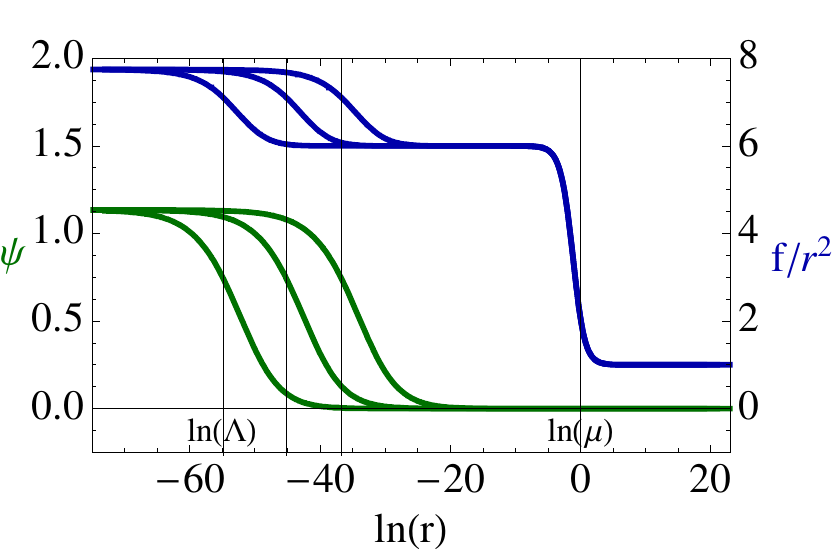}
\end{center}
\caption{\label{metricfns} Metric function $f/r^2$ (top curves) and
the scalar field $\psi$ (bottom curves) for different
values of $\kappa$ close to the critical point. The model
is the same as in Fig. \ref{fig:ab} with a fixed $g = -0.125$. The
two important scales are clear, the chemical potential $\mu$
which represents a cutoff on the critical $AdS_2$ theory
and $\Lambda = ( -\kappa_{\rm IR} )^{1/(1-2 \delta_-)}$
where $\delta_-$ is the critical IR dimension of operator
dual to $\psi$. 
} 
\end{figure}

First we study the metric functions of the ordered phase close to this point.
The goal here is to give a picture where it is clear that the bulk geometry
comes close to an intermediate critical point.
Fig.~\ref{metricfns} shows $f/r^2$ and $\psi$ for flows
that have $\kappa$ close to $\kappa_c$. Three separate regions are clearly visible in the figure. For small $r$, $\psi$ is nonzero and the solution approaches  $\wAdS_2 \times R^2$. At intermediate $r$, $\psi=0$ and the solution is  approximately $AdS_2 \times R^2$. This is the critical region. At large $r$, we approach $AdS_4$.\footnote{Similar
``3 shelf'' structure was seen in \cite{Faulkner:2008hm} for a probe brane
model. It would be interesting to understand better the critical theory involved in that case.}

\begin{figure}[h!]
\begin{center}
\includegraphics[scale=.95]{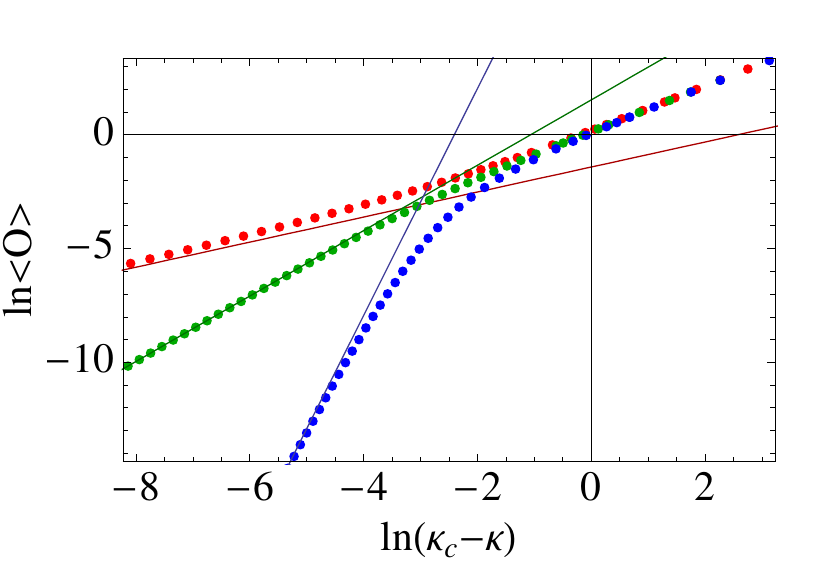}
\includegraphics[scale=.92]{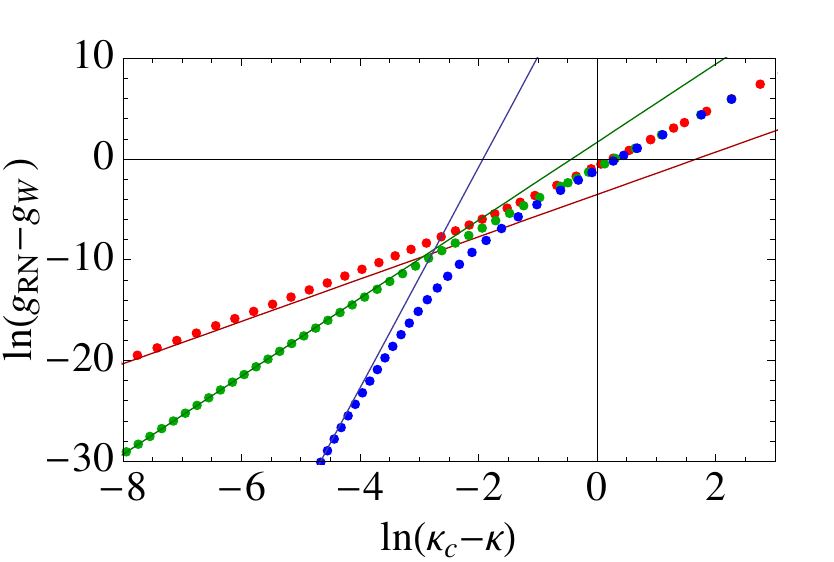}
\end{center}
\caption{\label{fig:ord}
The left figure shows the order parameter close to the critical point
for the three examples considered previously in Fig.~\ref{fig:ab}
(with the order of the curves flipped.)
For large negative $\kappa$ the data all follow the same
curve determined by the UV conformal dimension $\Delta_-$.
However for $\kappa \rightarrow \kappa_c$ the curves
diverge depending on their critical conformal dimension,
$\delta_- = .26, .37, .45$.
The right figure shows the free energy difference. 
}
\end{figure}

We then use precision numerics to plot the order
parameter and the free energy difference close to the critical
point. These are shown in Fig.~\ref{fig:ord} for different values of 
the scaling dimension $\delta_-$ given in (\ref{defdelta}). This scaling
dimension is distinct from ${\wDelta_-}$, and its importance
was made clear in Section 3. As Fig.~\ref{fig:ord} shows,  the data fits well to the  following scaling relations
to within $5\%$ error (this error could be reduced by going
closer to the critical point):
\begin{equation}
\label{scaling}
\left< \mathcal{O} \right> \propto (\kappa_c - \kappa )^{ \frac{\delta_-}{1 - 2 \delta_- }}
\,, \qquad g_W - g_{RN} \propto ( \kappa_c -\kappa)^{  \frac{1}{1-2 \delta_- }}
\end{equation}
The former was predicted in (\ref{eq:order}), and the latter can be derived from the former using (\ref{abint}).

Finally one point we would like to emphasize is
that at the end of the day the details of bulk coupling functions $V,G,J$ 
and the IR geometry $X$ will
not be important for understanding physics near 
the critical point.  Although, as already mentioned,
there are a few consistency conditions
which depend on the bulk coupling functions and are required in order
to realize this continuous second order transition. 
These are the conditions $s_c > 0$ defined in (\ref{defsc}) and $s_c^{\rm IR} >0$
defined in (\ref{defscir}). Also when it is important
we require $u_{\rm IR} > 0$ defined in (\ref{nonlin}). Although we have
not attempted to systematically compute these, as
we have shown above it is not
hard to find models in which these conditions are satisfied.

\subsection{Critical solution}

The size of the intermediate $AdS_2$ region grows without bound as $\kUV \rightarrow \kappa_c$ from below. At the critical point, the asymptotic $AdS_4$ part of the solution completely decouples from the $\wAdS_2$ IR region. There are two limiting solutions: One is the standard extreme RN AdS black hole which keeps the asymptotic $AdS_4$ region, and the other is a solution which asymptotically approaches $AdS_2\times R^2$ in the UV. 

Since we can compactify $R^2$ into $T^2$, this seems to contradict earlier arguments that there are no nontrivial solutions with asymptotically $AdS_2$ (times compact) boundary conditions \cite{Maldacena:1998uz}.  One way to see the potential problem is to consider a congruence of null geodesics at constant $\vec x$ in the metric (\ref{eq:ansatz}). Their convergence is  $c \propto \pm h'/h$. As long as $h$ is not constant everywhere,  either the left or right moving geodesics will have $c> 0 $ somewhere. However,  the Raychaudhuri equation and the null energy condition imply that if $c>0$ at one point, it must diverge in finite affine parameter causing a spacetime singularity. The resolution is that our limiting solution does have a singularity, but it is inside the Poincare horizon of the $\wAdS_2$ IR region. In fact, its Penrose diagram is similar to the extreme RN AdS solution. In short, the obstruction \cite{Maldacena:1998uz} applies only to solutions which are asymptotically $AdS_2$ in both the left and right asymptotic regions, and our critical solution has only one asymptotic region. 

We have focussed on a neutral scalar field in this section for simplicity. If $q \ne 0 $, then the natural IR geometry is  a new $\wAdS_4$ geometry with $\psi$ sitting at the global minimum of the potential. In this case, the critical solution flows from $AdS_2$ in the UV to $\wAdS_4$ in the IR, just the opposite of the standard RN AdS case.

\sect{Discussion}

In the growing literature on applications of gauge/gravity duality to condensed matter, one class of relevant operators has been largely ignored: multi-trace deformations. We have started to remedy the situation by studying various effects of these deformations. This includes spontaneous symmetry breaking and new quantum critical points. 
Below we first summarize the behavior near the quantum critical points, and then discuss various applications and generalizations of our results.


\subsection{Summary of critical exponents}

We have derived various scaling relations which hold near the quantum critical point which we now summarize. Let the operator dual to our bulk scalar, $\Op$, have dimension $\Delta_-$, and $\kappa$ be the coefficient of a double trace perturbation as in  (\ref{eq:dblUV}). Then for $\mu = 0$, the critical point is $\kappa = 0$ and near this point (from (\ref{neutralscaling}) and (\ref{k1})):
\be
 \langle \Op\rangle \propto \left(-{\kUV}\right)^{\Delta_-/(3-2\Delta_-)},\qquad
 T_c \propto (-\kUV)^{1/(3-2\Delta_-)}
\ee

For $\mu\ne 0$, the critical point is typically at a nonzero $\kappa_c$. For $\kappa$ close to $\kappa_c$ the bulk solution has a large intermediate $AdS_2\times R^2$ region. Let $\delta_-$ be the dimension of the operator dual to the bulk scalar in the corresponding $0+1$ dimensional CFT. There are a few cases depending on $\delta_-$. For $1/4 < \delta_- <1/2$, the critical point can be viewed as turning on a double trace operator with negative coefficient in the $0+1$ CFT. The exponents are given by (\ref{eq:order}) and (\ref{critemp1}) which are just the IR analogs of the ones above. Setting $\kIR \propto \kappa - \kappa_c$: 
\begin{equation}
\langle \mathcal{O}\rangle
\propto \left(-{\kIR} \right)^{\delta_-/(1-2\delta_-)}, \qquad 
 T_c \propto \left( - \kIR \right)^{1/(1 -2 \delta_-)}  \label{eq:NMF1}
 \ee
 For $0< \delta_- <1/4$, there are relevant higher multi-trace deformations. If the phase transition remains second order, then (\ref{quartic}) gives
 \begin{equation}
\langle \mathcal{O}\rangle
\propto \left(-{\kIR} \right)^{1/2}, \qquad 
 T_c \propto \left( - \kIR \right)^{1/(1 -2 \delta_-)}  
 \ee
 For  $\delta_-<0$, we have mean field behavior (from (\ref{mf5}) and (\ref{critemp2}))
 \be
\langle \mathcal{O}\rangle  \propto ( - \kIR)^{1/2} , \qquad T_c \propto  (- \kIR)
\ee
 Note that the exponent of $\langle \mathcal{O}\rangle$ is continuous at $\delta = 1/4$, and the exponent of $T_c$ is continuous at $\delta_- = 0$.

We have also found that at the critical point, $\kIR = 0$, there is a free gapless mode which satisfies $\omega \sim |\vec{p}\, |^{z}$ where the dynamical critical exponent $z$ is given by
\bea
 z&=&2  \qquad \qquad q\ne 0\ {\rm and}\ \delta_- < 0 \\
 z&=&1 \qquad \qquad q= 0\ {\rm and}\ \delta_- < -1/2 \\
 z &=& \frac{2}{1 -2\delta_- } \quad {\rm otherwise}\label{eq:NMF2}
\eea

\subsection{Applications of our results}

As we discussed in section two, one main application of our results (with $q\ne 0$) is to holographic superconductors. Since double trace deformations can break a $U(1)$ symmetry even with zero net charge density, they provide a new mechanism for constructing gravitational duals of superconductors. One appealing aspect of the new construction is that, despite translation invariance,  the DC conductivity in the normal phase is finite. We also showed that adding  a double trace deformation to the traditional construction of holographic superconductors introduces a sensitive knob for adjusting the critical temperature. 

Another possible application is to neutral 
order parameters (neutral under the $U(1)$ charge density). 
As mentioned in the introduction these can be used 
to model the onset of anti-ferromagnetic order \cite{Iqbal:2010eh}. 
Interestingly we found examples where we could
drive the critical temperature to zero, revealing a quantum
critical point characterized by the disappearance of
anti-ferromagnetic order. The nature of the critical point
is rather mysterious, however it's locally quantum nature discussed
around (\ref{univ}) has appeared in previous theoretical studies
of heavy fermion criticality \cite{si1,si2}. Indeed measurements \cite{exp} 
of the dynamic susceptibility of a certain heavy fermion material $CeCu_{6-x} Au_x$ close to criticality 
seem  consistent with the form of the two point function that we derive.
Optimistically our results based on $AdS_2 \times R^2$ might
shed some light on the robustness of the large dimension limit 
used to justify the theoretical results of \cite{si1,si2}. Or turning this the other
way, we might learn something about the mysterious CFT dual to $AdS_2 \times R^2$. 
Such possibilities were discussed recently in a related context \cite{Sachdev:2010um}.
The relationship of $AdS_2$ to local quantum criticality was discussed
in \cite{Si}.

We now would like to discuss
some immediate generalizations.

\subsection{Magnetic fields}
Studying the theory when a magnetic field is turned on is often difficult, due to the formation of superconducting droplets at nonzero $q$. However, we can study the theory near the critical point following \cite{Hartnoll:2009sz, Horowitz:2010gk} by studying linearized analysis around a dyonic black hole. The dyonic RN AdS black hole\footnote{See \cite{Hartnoll:2007ip} for a detailed study of the transport properties of the dyonic black hole.} is a simple generalization of (\ref{rnbh}), which now we are writing in terms of the horizon radius $r_0$, the chemical potential $\mu=\rho/r_0$, and the magnetization $M=B/r_0$ (not to confused with the mass of the scalar field $m$)
 \be
ds^2=  -f dt^2 + r^2 (dx^2+dy^2)+\frac{dr^2}{f}\,, \quad A=\mu\left(1 -\frac{r_0}{r}\right)dt+M r_0 x dy,
\ee
\be
f = r^2 -\frac{m_0}{2 r }+\frac{(\mu^2+M^2)r_0^2}{4r^2}\,,\quad m_0 = \frac{\left( \mu^2+M^2\right)r_0}{2}+2 r_0^3 \,,
\quad T = \frac{3 r_0}{4\pi}-\frac{\mu^2+M^2}{16\pi r_0}.
\ee
We can generalize our calculations of the critical temperature by studying linearized fluctuations, a simple generalization of  \cite{Hartnoll:2009sz, Horowitz:2010gk, Iqbal:2010eh}. 

For now let us restrict to $q=0$, and study the effect of a nontrivial $G$. The important equation is the generalization of (\ref{psieq}), with $\Psi=R(r) e^{iky-i\omega t}$, which leads to the wave equation
\be
(r^2fR')'+\left[\frac{(r\omega)^2}{f}+\frac{g(\mu^2-M^2)r_0^2}{2r^2} -m^2r^2\right]R=0.\label{Reom}
\ee
This can be solved either numerically or by a matched asymptotic expansion. Since we are working at $q=0$, turning on a magnetic field effectively shifts $g$ for our linearized fluctuations. If we restrict to zero temperature and frequency, we  find that the effective $AdS_2$ mass $m_{eff}^2$ changes as a function of $M$,
 \be
 m_{eff}^2=\frac{m^2}{6}+g\left(\frac{M^2-\mu^2}{M^2+\mu^2}\right),\label{mag_mass}
 \ee
 where again $\delta_\pm=\frac{1}{2}\pm\sqrt{\frac{1}{4}+m_{eff}^2}$ is the infrared dimension. It is clear that turning on a magnetic field changes the infrared dimension of the operator, and can possibly change it from real to complex\footnote{For $q\neq 0$ this was pointed out in \cite{Iqbal:2010eh}, and we see that a nontrivial $G$ can cause this to occur for $q=0$ as well. The magnetic field at which this occurs when $G=1,~q\neq 0$ is the $B_c$ found in \cite{Hartnoll:2008kx}.}. This defines a critical magnetization where $m_{eff}^2=-1/4$, 
 \be
M_c=\mu \sqrt{\frac{12g-2m^2-3}{12g+2m^2+3}},
 \ee
 where there is an infinite order holographic BKT transition
 \footnote{For a different type of quantum phase transition that
 occurs as a function of a magnetic field, see \cite{D'Hoker:2010rz}.}, just as that found in \cite{Jensen:2010ga,Iqbal:2010eh} and studied further in \cite{Jensen:2010vx,Evans:2010hi,Pal:2010gj} (for
 a field theoretic discussion see \cite{Kaplan:2009kr}.)
 The critical point at $\kappa=\kappa_c$ turns into a second order phase boundary, which terminates at $M_c$. See figure \ref{fig:mag_phase} for example phase diagrams. 
   \begin{figure}[h!]
\begin{center}
\includegraphics[scale=.75]{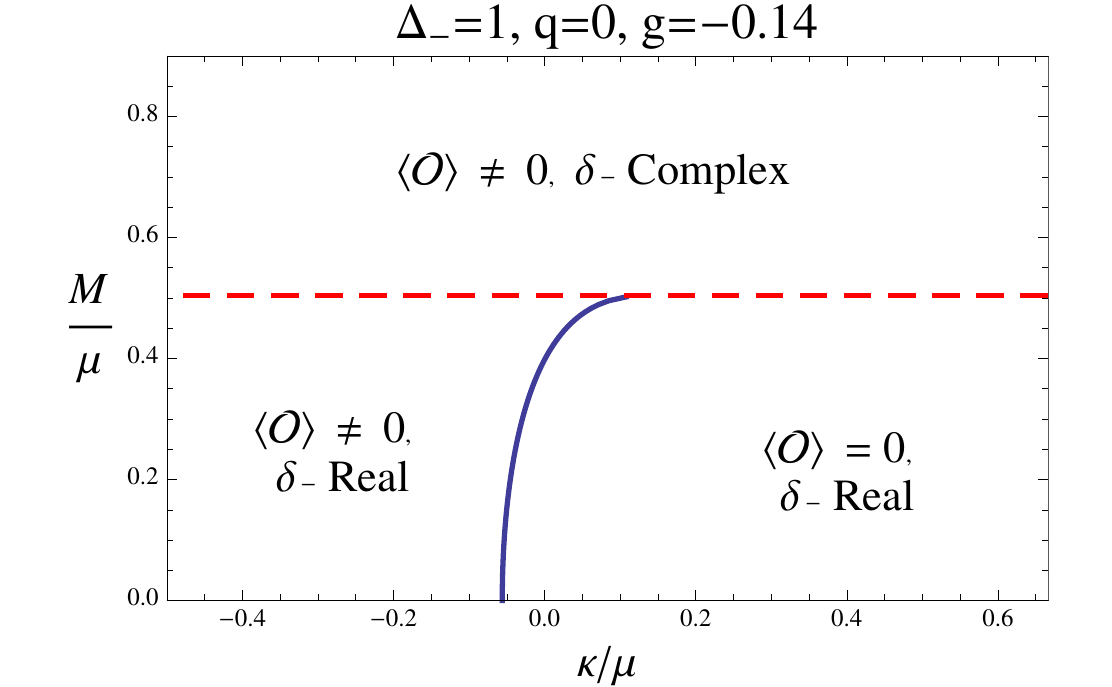}\includegraphics[scale=.75]{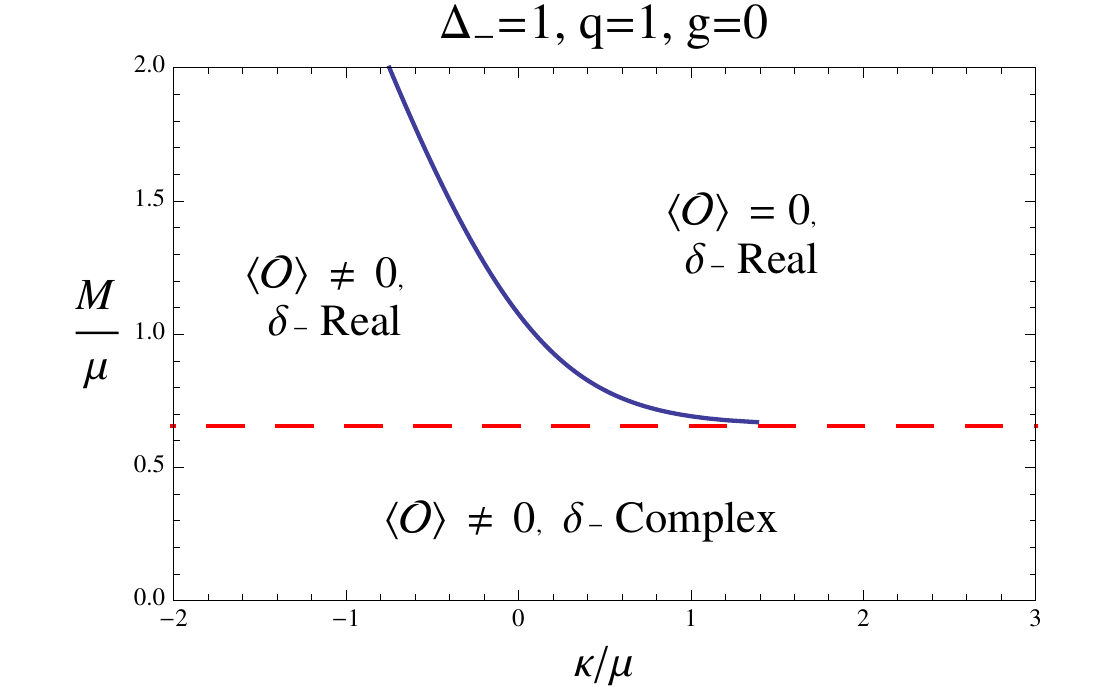}
\end{center}
\vspace{-.5cm}
\caption{\label{fig:mag_phase}
Phase diagrams when turning on a magnetic field. On the left, when $g$ is negative and $q$ is zero, turning on a magnetic field helps destabilize the disordered phase. On the right, when $q$ is nonzero and $g$ is zero, turning on a magnetic field helps stabilize the disordered phase. The red dashed line is the holgraphic BKT transition at $M=M_c$, an infinite order transition. The solid blue line is the non-mean field transition controlled by $\kappa_{IR}$ changing sign, where $\delta_-\lesssim1/2$. At the holographic BKT line the critical exponents diverge.}
\end{figure}

 Along our $\kappa_c$ transition line, moving towards the BKT transition, $\delta_-\rightarrow 1/2$ which implies the critical exponents of the transition (\ref{eq:NMF1}, \ref{eq:NMF2}) diverge.  This must match onto the known holographic BKT behavior \cite{Iqbal:2010eh}, where
\be
T_c,\langle\Op\rangle\sim \exp\left[-\pi/\sqrt{-3/2-6m_{eff}^2}\right].
\ee
It would be interesting to understand the behavior of the order parameter two point function as we approach the holographic BKT transition, and how it matches on to the known behavior along the second order transition line. At nonzero $q$, (\ref{mag_mass}) generalizes to
  \be
m_{eff}^2=\frac{m^2}{6}-\frac{q^2\mu^2}{3(M^2+\mu^2)}+\frac{qM}{\sqrt{3(M^2+\mu^2)}}+g\left(\frac{M^2-\mu^2}{M^2+\mu^2}\right).
 \ee

 \subsection{Quantum corrections}

We will now discuss the effect of quantum corrections 
governed by an expansion in the gravitational coupling $G_N$.
These correspond to $1/N$ corrections in the field theory. 
We will be particularly interested in how
robust our results for the critical exponents are to bulk quantum corrections.

We start by considering the disordered phase. Firstly
it would be interesting to understand one loop
corrections to the thermodynamics and transport
that come from fluctuations of $\psi$. Methods
developed in 
\cite{Denef:2009kn,Denef:2009yy, Hartnoll:2009kk, Hartman:2010fk, Faulkner:2010da, Anninos:2010sq} 
would be useful for this purpose.
In the large $N$ limit, quite often bulk geometric contributions to thermodynamics
and transport dominate since they are of order $1/G_N$,
so 1 loop contributions which are of order $1$ in the $G_N$ expansion will
be suppressed.  However the mode that becomes gapless at the critical
point will give large contributions that may be isolated from the
geometric contributions due to strong IR behavior. This philosophy
was used in \cite{Denef:2009kn,Denef:2009yy, Hartman:2010fk} and in 
\cite{Faulkner:2010da} to extract thermodynamics (most notably quantum
oscillations) and charge transport of holographic non-Fermi liquids.

It is also important to understand
how these modes renormalize bulk couplings
when running in loops. Although this would require
a fairly complicated bulk one loop calculation, we can
argue what the outcome should be, based on the
low energy semi-holographic model that 
we wrote down in (\ref{wf}). Note that in the disordered
phase this model reproduces the 2 point function
$\chi_R$ for all $\delta_-$ real. 
Consider here the neutral case $q=0$, then
ignoring the coupling $\eta$ between the boundary
mode $\Phi$ and the CFT operator $\Psi$ what
we wrote done in  (\ref{wf}) was simply a
$\Phi^4$ theory in $2+1$ dimensions. Thus
we  expect large quantum corrections
to mean field at low energies, since $u_\Phi$
should run to strong coupling 
(at the Wilson Fisher fixed point). Low
energies should be $\omega \sim G_N$ in this case.

We believe the real situation when $\eta$ is non-zero
will be slightly different since the modes in the 2 point function for $\Phi$
now have a different
critical exponent $z = \rm{max} (1, 2/(1-2\delta_-) )$.\footnote{
This discussion was inspired by similar considerations
in \cite{si2}. } This changes
the RG power counting of the non-linear coupling $\Phi^4$ effectively
softening the running. One can express this in terms
of an effective dimensionality
$d_{\rm eff} = 2 + z$, where for $d_{\rm eff}>4$ 
one is above the upper critical
dimension of $\Phi^4$ theory and  non-linear effects should not be important
(except in the case that they are dangerously irrelevant, which
here they are not.) 
Thus the scaling
exponents we derived should not receive corrections
as long as $z>2$. 
Interestingly when $z<2$
we should be able to see a Wilson Fisher type fixed point
as an expansion in $\delta_-$, analogous to the classic $d-4$ expansion 
since $d_{\rm eff} \approx 4+ 4 \delta_-$ for small $\delta_- <0$.
This will then change the mean field predictions we found above,
for example in (\ref{mf5}).  
We leave details for future investigation.
This softening of nonlinear interactions
occurs in a similar context for holographic non-fermi liquids,
leading to suppression of the BCS instability \cite{Hartman:2010fk}.

On the ordered side, of course the second order
mean field transition that occurs along the line $T=T_c (\kappa)$
will not survive corrections beyond mean field theory.
In $2+1$ dimensions at finite temperature fluctuations
of a continuous order parameter destroy long range order.
This is a consequence of the Coleman-Mermin-Wagner theorem
studied in a holographic context in \cite{Anninos:2010sq}. 
Ultimately on very long distance scales ($\sim e^{ \# /G_N}$) this 
mean field transition should be replaced by a (conventional)
BKT transition. 
It would be very interesting to understand how these corrections scale 
close to the critical point. At zero temperature
since fluctuations in the temporal direction become
important it is now possible to have continuous order. (One
should not be fooled by the fact that the critical theory
seems to be controlled by a $0+1$ dimensional theory.
The IR of the zero temperature ordered phase has
nothing to do with this $0+1$ dimensional theory, and
besides from the bulk perspective the $R^2$ directions of $AdS_2 \times R^2$ are infinite and still
allow for fluctuations.)

\subsection{Lifshitz normal phase}

One possible generalization of the critical point story is to replace
the normal phase, which for us was the RN black hole,
with a more general charged black hole solution, which
is however still asymptotically $AdS_4$. There are 
several motivations for doing this, one is
that this  may solve the finite entropy problem at extremality,
that plagues the RN black hole. In this vein one 
expects quantum corrections (orthogonal to the
ones discussed above) to lift the finite degeneracy
associated with this finite entropy. In so doing, these
corrections must replace the $AdS_2 \times R^2$ critical
solution with something else.
This is exactly what happens when one studies the back reaction
of fermions in the bulk \cite{Hartnoll:2009ns}, where the IR $AdS_2 \times R^2$ solution
is replaced by a Lifshitz geometry \cite{Kachru:2008yh} with a dynamical
critical exponent $z_G \sim \mathcal{O}(1/G_N)$ (we use a subscript
$G$ to denote a geometric exponent, to be contrasted with our $z$.)
For $z_G \rightarrow \infty$ one recovers the $AdS_2 \times R^2$
geometry. So finite $z_G$ is playing the role of a regulator 
and we expect this to be a general phenomenon. 
Note for example now the entropy scales as $S \sim T^{2/z_{G}}$
at low temperatures \cite{Goldstein:2009cv}.

Another motivation is simply to have a more general
understanding of what holographic models can
do at finite density. One simple generalization is
to add a dilaton field $\sigma$ which couples to the gauge field, considered
first in the AdS/CMT context in \cite{Goldstein:2009cv}.\footnote{See \cite{Charmousis:2010zz}
for further generalizations. } We will consider this model here
as a concrete example. In our action (\ref{action}) one
replaces $F^2 \rightarrow e^{2 \alpha  \sigma} F^2$ and adds
a kinetic term for $\sigma$ with normalization $ - 2 (\nabla \sigma)^2$ . We will not add a potential for $\sigma$.
There are many choices for how to couple $\sigma$ to $\psi$, 
we begin for simplicity by assuming no extra couplings between $\sigma$ and $\psi$
other than through the replacement above. 
However we will argue that other choices of couplings should
produce similar results.

The normal phase has $\psi=0$. The appropriate solution
which is charged and asymptotically $AdS_4$ in the UV 
was identified in \cite{Goldstein:2009cv}. It
has a  Lifshitz like geometry in the IR,
replacing the $AdS_2 \times R^2$ solution of the RN black hole. Roughly
speaking the dilaton runs $\sigma \rightarrow \infty$ such that the coupling
and the 
charge density redshift away in the IR. So there
is no need for a charge on the horizon to source the electric field.
The critical exponent  is determined by the dilatonic coupling
$z_G = 1+ 2/\alpha^2$.  The IR scaling geometry 
at finite temperature is,
\bea
\label{finitetz}
ds^2 &=& - \hat{f} dt^2 + \frac{ d\hat{r}^2}{\hat{f}} + \hat{r}^{2/z_G} d\vec{x}^2
\,, \quad \hat{f} = \mathcal{N} \hat{r}^2 \left(1 - (\hat{r}_T/\hat{r} )^{1+2/z_G} \right) \\
\sigma &=& - \frac{\sqrt{z_G-1}}{z_G} \log (\hat{r}/\mu) \,, \quad
\quad A = \mathcal{M} \mu^{-2/z_G} \left( \hat{r}^{1+2/z_G} - \hat{r}_T^{1+2/z_G} \right)dt
\eea
where $\mathcal{N}, \mathcal{M}$ are numbers determined by $z$
and $\hat{r}_T = (2/3) \pi (1+1/z_G) T$. This solution
should be compared to (\ref{finitet}).
Note that
the solution is not precisely scale invariant due to the form of both
the gauge field and the dilaton, to emphasize
this we have included explicit factors of $\mu$.

In this background we can now repeat our construction
of the critical point. We will only consider the disordered phase,
where the 2 point function determines most of the physics.
The linear fluctuations of $\psi$ on (\ref{finitetz}) have a scale invariant form 
for $\hat{r} \ll \mu$, with dimensions $[\omega] = z_G [\vec{p}] = [T] =  [\hat{r}]$.
This is a nontrivial statement because the background is
not scale invariant, so in the limit $\hat{r} \ll \mu$ various
couplings disappear from the equation for $\psi$. 
For example since the gauge potential
$q \phi$ redshifts away compared to the energy $\omega$, 
the coupling $q$ will not play a role in IR physics.
Including more general couplings between $\sigma$ and $\psi$
will result in different IR scaling limits;  the only
assumption one needs in order to extend these results, is that
this limit exists (and is nontrivial.)\footnote{ 
One may also need to generalize the scaling  to 
$[\omega] = z_G [\vec p] = n [\hat{T}] =  n [\hat{r}]$ for some $n$.}
  
Imposing incoming boundary conditions on the solution to this scale invariant differential equation,
for large $\hat{r}$ one finds a generalization of (\ref{sigmaR}).
\be
\psi(\hat{r}) \sim \left(  \hat{r}^{-\delta_-} + \Sigma_R(\omega,T,\vec{p}) \right) 
= \left(  \hat{r}^{-\delta_-} +  T^{\delta_+ - \delta_-}
 g \left(\omega/T , \vec{p}^{\,\,2}/T^{2/z_G} \right)  \,
\hat{r}^{-\delta_+} \right)
\ee 
where $\Sigma_R$ is the IR retarded Green's function,
which now has general nonanalytic dependence on $\vec{p}^2$.
The critical conformal dimensions are given by,
\be
\delta_\pm =  \frac{ d_{ \rm eff}^G}{2z_G}
\pm \sqrt{ m_{\rm eff}^2 + \left(\frac{ d_{\rm eff}^G}{2z_G}\right)^2 }
\,, \quad m_{\rm eff}^2 = \frac{d_{\rm eff}^G}{z_G} \left( \frac{ \Delta (\Delta-3)}{6} 
\left( 1+ \frac{1}{z_G} \right) - g \left( 1 - \frac{1}{z_G} \right) \right)
\ee
where $d_{\rm eff}^G = 2 + z_G$. 
Note in particular any dependence on $q$ has vanished. We do not however 
expect the expression for $m_{\rm eff}$ to be universal
when considering other couplings of $\psi$ to the dilaton.
Matching to the outer region we find for the IR limit
of the two point function,
\be
\label{univ2}
\chi_R =  \frac{Z}{\kIR + c_p \vec{p}^2 - T^{2/z} g(\omega/T, \vec{p}^2/T^{2/z_G} ) }
\ee
where $z = 2/(\delta_+ - \delta_-)$. Note that
this result can also be obtained semi-holographically \cite{Faulkner:2010tq}.
Again there should be some matching between $\kIR$
and $ (\kappa - \kappa_c)$ in the outer region.
Importantly we have included an analytic in $\vec{p}^2$ correction
which will also come from the outer region. This correction
will be important if $z < z_G$ as we discuss now.

The coherent
part of the two point function comes from the dispersing
mode with $\vec{p}^2 \sim T^{2/z} \sim \omega^{2/z}$. 
Along this mode the second argument in the scaling function
$g$ vanishes, $\vec{p}^2/T^{2/z_G}   \rightarrow 0$.
That is, we can ignore the momentum dependence
in the scaling function $g$. Thus we are forced
back to a locally quantum critical theory, where the 2 point
function is effectively analytic in $\vec{p}$.  The critical exponent $z$
is still determined by the conformal dimension of the operator
dual to $\psi$ in the Lifshitz IR theory.  

Of course this analyticity in momentum and the critical exponent $z$
will not persist away from the coherent peak, and the 2 point
function will have signatures of the $z_G$ Lifshitz scaling
for $ \vec{p}^2 \sim \omega^{2/z_G}$.
These signatures will be incoherent and subdominant. They
are analogous to the gapless modes that are present
in the $AdS_2$ case, and also present in the holographic
non-Fermi liquids \cite{Faulkner:2009wj}. This
makes sense in the context of the limit $z_G \rightarrow \infty$
where one recovers $AdS_2$.

For $z>z_G$ then the analytic $\vec{p}^2$ term
is subdominant and the two point function
will have strong nonanalyticity in both frequency
and momentum. The critical point is then truly
governed by the Lifshitz geometry.

So to summarize, for the case $z< z_G$ we see
that our results for the $AdS_2$ critical point
are robust under generalization
to a Lifshitz IR scaling geometry.  Further it is easy to see that the phase
boundary is still controlled 
by the conformal dimension $T_c \sim ( -\kIR)^{1/(\delta_+- \delta_-)}$ 
and the order parameter turns on as $\left< \mathcal{O} \right> 
\sim   ( -\kIR)^{\delta_-/(\delta_+-\delta_-)}$
which is similar to what we found earlier, although now we have $\delta_+ + \delta_- = d_{\rm eff}^G/z_G$. A more thorough discussion is in order,
but we leave this to future work.

 


\vskip 1cm
\centerline{\bf Acknowledgements}
\vskip .5 cm
It is a pleasure to thank Hong Liu, Don Marolf, and Joe Polchinski for discussions. TF
would also like to thank Kristan Jensen and Andreas Karch for colloboration on related
matters. This work was supported in part by the US National Science Foundation under Grant No.~PHY08-55415, Grant No. PHY05-51164 and the UCSB Physics Department.

\appendix

\sect{Equations of motion}\label{appendix:EOM}
\subsection{Field equations}
We begin with the action
\be\label{appaction}
S = \int d^4x \sqrt{-g} \left( R- \frac{1}{4} G(\psi) F^2 - (\nabla \psi)^2-J(\psi)(\nabla \theta - q A )^2 - V(\psi)\right),\ee
where we have decomposed a complex scalar $\Psi= \psi e^{i\theta}$. We have normalized $A$ such that 
\be\label{Gdef}
G=1+g \psi ^2+\Op(\psi ^4),\ee
and periodicity of $\theta \sim \theta+2 \pi$ implies
\be
J= \psi ^2+\Op(\psi ^4).
\ee
Lastly we assume 
\be
V = -6+m^2 \psi ^2+\Op(\psi ^4),
\ee
where we have chosen units where the AdS radius is unity. The wave equations for the scalar are
\be
\nabla^2 \psi-\frac{1}{2}V'(\psi)-\frac{1}{8}G'(\psi)F^2-\frac{1}{2}J'(\psi)(\nabla \theta-q A)^2=0,\label{modulusEOM}
\ee
\be
\nabla_\mu\left[ J(\psi) (\nabla^\mu\theta - q A^\mu)\right]=0.\label{phaseEOM}
\ee
Maxwell's equation is
\be
\nabla_\mu [G(\psi)F^{\mu\nu}]+2 q  J(\psi)(\nabla^\nu\theta-qA^\nu)=0.
\label{maxwellEOM}
\ee
Einstein's equation is
\bea
R_{\mu\nu}-\frac{1}{2}g_{\mu\nu}\left(R-\frac{1}{4}G(\psi)F^2-(\nabla \psi)^2-J(\psi)(\nabla\theta-q A)^2-V(\psi) \right)\nonumber
\\
-\frac{1}{2}G(\psi)F_{\mu\rho}F_\nu{}^\rho-\nabla_\mu \psi\nabla_\nu \psi-J(\psi)(\nabla_\mu\theta-qA_\mu)(\nabla_\nu\theta-qA_\nu)=0.\label{einsteinEOM}
\eea

\subsection{Our ansatz}

When studying static black hole solutions, we make the ansatz
\bea
ds^2&=&-f(r)dt^2+dr^2/f(r)+h(r)^2(dx^2+dy^2),\nonumber
\\
A&=&\phi(r)dt, \quad \psi = \psi(r), \quad \theta=0.
\eea
This leads to the equations of motion
\be
\psi''+\left(\frac{f'}{f}+\frac{2h'}{h} \right) \psi'-\frac{V'(\psi)}{2f}+\frac{G'(\psi)\phi'^2}{4f}+\frac{q^2\phi^2J'(\psi)}{2f^2}=0,\label{eq:1}
\ee
\be
\phi''+\left(\frac{2h'}{h}+\frac{G'(\psi) \psi'}{G(\psi)}\right) \phi'-\frac{2q^2J(\psi)}{G(\psi)f}\phi=0,\label{eq:2}
\ee
\be
h''+\frac{\psi'^2}{2}h+\frac{q^2J(\psi)\phi^2}{2f^2}h=0,\label{eq:3}
\ee
\be
\frac{h'^2}{h}+\frac{f'h'}{f}-\frac{h\psi'^2}{2}+\frac{hV(\psi)}{2f}+\frac{G(\psi)h\phi'^2}{4f}-\frac{q^2hJ(\psi)\phi^2}{2f^2}=0.\label{eq:4}
\ee
There is a third component of Einstein's equations which is a dynamical equation for $f$ and can be derived from the above equations,
\be
f''+\frac{2f'h'}{h}+V(\psi)-\frac{G(\psi)\phi'^2}{2}-\frac{2q^2J(\psi)\phi^2}{f}=0.\label{eqExtra}
\ee 
The equations of motion are invariant under the following symmetries, which leave invariant the scalar field, the line element, and the Maxwell one-form: First, a shift symmetry,
\be
r\rightarrow r+a.
\ee 
Second, a conformal rescaling,
\be
r\rightarrow \Lambda r,~ (t,x_i)\rightarrow (t,x_i)/\Lambda,~f\rightarrow \Lambda^2 f,~ h\rightarrow \Lambda h,~ \phi \rightarrow \Lambda \phi.
\ee
Lastly, a spacial rescaling
\be
x_i\rightarrow x_i /s,~ h\rightarrow s h.
\ee

\subsection{Calculating conductivity}

To calculate conductivity in this background, we study linearized modes 
\be
\delta A_x = a_x(r)e^{-i\omega t}, \quad \delta g_{tx}=g_{tx}(r)e^{-i\omega t}.\ee
These have the following equations of motion
\bea
a_x''&+&\left(\frac{f'}{f}+\frac{G'(\psi)\psi'}{G(\psi)} \right)a_x'+\left(\frac{\omega^2}{f^2}-\frac{2q^2J(\psi)}{fG(\psi)} \right)a_x+\frac{\phi'}{f}g_{tx}'-\frac{2h'\phi'}{fh}g_{tx}=0,\nonumber
\\
g_{tx}'&-&\frac{2h'}{h}g_{tx}+G(\psi)\phi' a_x=0.
\eea
We can eliminate $g_{tx}$ and find
\be\label{origcond}
a_x''+\left(\frac{f'}{f}+\frac{G'(\psi)\psi'}{G(\psi)} \right)a_x'+\left(\frac{\omega^2}{f^2}-\frac{G(\psi)\phi'^2}{f}-\frac{2q^2J(\psi)}{fG(\psi)} \right)a_x=0.
\ee
There is now a factor of $G(\psi)$ multiplying the Maxwell term in the action, but since $\psi =\alpha/r^{\Delta_-}+\ldots$ and $\Delta_->1/2$, $G$ goes to one rapidly enough at the boundary that the analysis of \cite{Hartnoll:2008kx} follows and we can read off the conductivity from the asymptotics of $a_x$,
\be
a_x=a_x^{(0)}+a_x^{(1)}/r+\ldots,~\sigma = -ia_x^{(1)}/\omega a_x^{(0)}.
\ee
As in \cite{Horowitz:2009ij} we can rewrite this as Schr\"odinger's equation. Using a new radial variable and rescaling the gauge field
\be
dz = \frac{dr}{f}, \quad b=\sqrt{G(\psi)}a_x,
\ee
we find that (\ref{origcond}) reduces to
\be
-b''(z)+V_{Sch}(z)b(z)=\omega^2b,\label{app:schrodinger}
\ee
where
\be\label{schrodpot}
V(z) =f \left(  G(\psi)\phi_{,r}^2+ \frac{2q^2J(\psi)}{G(\psi)}+\frac{1}{\sqrt{G(\psi)}}\left[\frac{f G'(\psi)\psi_{,r}}{2\sqrt{G(\psi)}}\right]_{,r}\right).
\ee
We can relate conductivity to the reflection amplitude. In the new coordinates, near the boundary at $z=0$, where $z=-1/r+\ldots$, we assume $b = e^{-i\omega z}+\R e^{i\omega z}$ and near the horizon at $z=-\infty$ ingoing boundary conditions require $b=\mathcal{T} e^{-i\omega z}$. We find
\be
\sigma =\frac{1-\R}{1+\R}.
\ee
We once again see that the necessary condition to find a hard gap in the real conductivity is that $V(z)$ not vanish on the horizon. In \cite{Horowitz:2009ij} it was proven that for $J=\psi^2, G=1$ the Schr\"odinger potential must vanish on the horizon. It is an interesting question as to whether one can evade this theorem with more general $J$ and $G$.

\sect{Critical temperature at zero chemical potential}\label{SSBTc}
We can find the critical temperature by looking for static linearized scalar hair on AdS satisfying our linear boundary conditions. In the background of a $d+1$ dimensional AdS-Schwarzschild black hole, the wave equation is
\be
\psi''+\left(\frac{f'}{f}+\frac{d-1}{r} \right) \psi'-\frac{\Delta(\Delta-d)}{f} \psi +\left(\frac{\omega^2}{f^2}-\frac{k^2}{r^2 f}\right) \psi =0,
\ee
where $f=r^2[1-(r_0/r)^d]$ and $T=dr_0/4\pi$. We have dropped the subscript and are working with $\Delta = \Delta_-$. The static case $k=\omega=0$ can be solved via hypergeometric functions. Writing $z=r_0/r$, 
\be
\psi = c_1 z^\Delta{}_2F_1[\Delta/d,\Delta/d,2\Delta/d;z^d]+c_2 (\Delta \leftrightarrow d-\Delta)
\ee
which at large $r$ falls off as
\be
\psi = c_1(r_0/r)^\Delta+c_2(r_0/r)^{d-\Delta}+\ldots
\ee
and therefore satisfies the linear boundary condition
\be
\frac{\beta}{\alpha}=\kUV=\frac{c_2}{c_1}(4\pi T/d)^{d-2\Delta}
\ee

Now recalling that ${}_2F_1[a,b,a+b;y] \propto \log(1-y)$ we must choose the appropriate linear combination to cancel the logarithmic term for regularity on the horizon (this logarithmic branch is coming from the $\omega=0$ limit of ingoing and outgoing waves on the horizon.) The appropriate combination is
\be
\frac{c_2}{c_1}=-\frac{\Gamma(2\Delta/d)\Gamma(1-\Delta/d)^2}{\Gamma(2-2\Delta/d)\Gamma(\Delta/d)^2}
\ee
This tells us that there is a second order phase transition as we heat the system up, and above the critical temperature
\be
T_c =\frac{d}{4\pi} \left(-\frac{\Gamma(\Delta/d)^2\Gamma(1-2\Delta/d)}{\Gamma(1-\Delta/d)^2\Gamma(-1+2\Delta/d)} \right)^{1/(d-2\Delta)}(-\kUV)^{1/(d-2\Delta)}
\ee
the system returns to the symmetry preserving state.

\sect{More on the 2 point function}

Here we give details of the computation of the two point function
in the ordered phase that we left out from Section 3.
Most importantly we include a discussion of finite but
small temperature, and compute the quantity $\mathbb{X}$ 
given in (\ref{X}).

The small frequency limit can be supplemented 
with a small $T$ limit, where $\omega/T$ is held fixed.
To do this, as in (\ref{small}), we redefine,
\be
\omega \rightarrow \epsilon \omega, \quad T \rightarrow \epsilon T,
\quad \vec{p} \rightarrow \epsilon \vec{p}
\ee
and proceed to expand in $\epsilon$ (setting $\epsilon=1$ at the end.)
The inner region now has the metric
of the $AdS_2$ black hole: 
\begin{equation}
\label{finitet}
ds^2_0 = - \hat{f} d\hat{t}^2 + \frac{ d\hat{r}^2}{\hat{f}} + d\vec{\hat{x}}^2
\,, \quad \hat{f} = 6 \hat{r}(\hat{r} - (2/3) \pi T ) \,,
\quad A_0 = \sqrt{12} (\hat{r} - (2/3) \pi T) d\hat{t}
\end{equation}
In this background we can compute the $AdS_2$ Green's
function for $\psi$, imposing incoming boundary conditions at the horizon.
Asymptotically at the $AdS_2$ boundary one has,
\be
\label{sigmaR}
\psi_0(\hat{r}) = \mathcal{N} \left(  \hat{r}^{-\delta_-} + \Sigma_R(\omega,T) \hat{r}^{-\delta_+} \right)
\ee

The answer is found in Appendix D Eq.~(27) of \cite{Faulkner:2009wj}.
It is reproduced here in (\ref{scaleform}) using our current conventions.
Higher order corrections to the inner region will never
be important, so we do not consider them.

In the outer region we expand $\psi(r) = \psi_0(r) + \epsilon \psi_1(r)
+ \epsilon^2 \psi_2(r) + \ldots$. Where:
\be
\label{st}
D \psi_0 = 0 \,, \quad D \psi_1 = X_1(\psi_0) \,, \quad 
D \psi_2 = X_2(\psi_0,\psi_1) \, \ldots
\ee
Where at zeroth order we have,
\be
D = \partial_r ( r^2 f_0 \partial_r ) - m^2 r^2 + \frac{g \rho_0^2}{2r^2} + q^2 r^2 \phi_0^2 /f_0
\,,\quad \phi_0 = \mu -\frac{\rho_0}{r}\,, \quad f_0 = r^2 -\frac{\rho_0 \mu}{3 r }+\frac{\rho_0^2}{4r^2} 
\ee
where we recall that $\rho_0 = \mu^2/\sqrt{12}$.  
And for the next two orders the source terms are:
\bea
X_1(\psi_0) &=& - \frac{T}{\mu} \frac{ 2 \pi}{\sqrt{3}} D( r \psi_0' )
+ 2 q \left(- \omega + \frac{2 \pi}{\sqrt{3}} T q \right) \frac{\phi_0 r^2}{f_0} \psi_0 \\
X_2(\psi_0,\psi_1) &=& \left( \vec{p}^2  - \frac{\omega^2 r^2}{f_0} \right) \psi_0 + \mathcal{O} (\omega ,
T ) \psi_1
\eea
where in $X_2$ we have ignored nonleading corrections. These
corrections will be of order $q \omega^2, q \omega T, T^2$, 
which are sub-leading compared to $X_1$. Note  that we have
kept an $\omega^2$ term in $X_2$ since for the neutral
case this is the leading correction in $\omega$. Given
this we can now treat $\psi_1 + \epsilon \psi_2 \approx \psi_L$ together ($L$ stands for leading corrections). So we now write,
\be
\label{towork}
D\left( \psi_L + \frac{2\pi T}{\sqrt{3} \mu} r \psi_0' \right)
= X_L \psi_0 \,, \quad 
X_L =  2 q \left(- \omega + \frac{2 \pi}{\sqrt{3}} T q \right)
\frac{\phi_0 r^2}{f_0} + \epsilon \left( \vec{p}^2  - \frac{\omega^2 r^2}{f_0} \right) 
\ee

The horizon and boundary behavior of $\psi_0$ was given in 
(\ref{exp0}).  Since we can always add  a homogenous solution to
$\psi_L$ we demand that at the horizon $\psi_L$ has
the following behavior

\bea
\label{lead}
\psi_L &=& B_2 (r-r_\star)^{-\delta_+ -2} + B_1 (r-r_\star)^{-\delta_+ -1} 
 + 0 \cdot (r-r_\star)^{-\delta_+}  \\ &+&  A_2 (r-r_\star)^{-\delta_- -2} + A_1 (r-r_\star)^{-\delta_- -1} 
 + 0 \cdot (r-r_\star)^{-\delta_-}
+ \ldots
\eea 
where $A_{1,2}$ and $B_{1,2}$ are completely determined by $\psi_0$ through
the source terms in (\ref{st}). The nontrivial requirement
relates to the terms we have set to zero above. Given this constraint
we can then match to the inner region via (\ref{sigmaR})
and (\ref{exp0}), we find $\hat{\beta}_0/\hat{\alpha}_0 = \Sigma_R$
with no contribution from $\psi_L$ (and here we have set $\epsilon=1$.)
Note the diverging terms in (\ref{lead}) can be matched
to lower order $1/\hat{r}$ corrections to (\ref{sigmaR}). 

Asymptotically
at the boundary of $AdS_4$ we have,
\bea
\psi_L = \alpha_L r^{-\Delta_-} + \beta_L r^{- \Delta_+}
\eea
and from this and (\ref{exp0}) we can read off
the 2 point function:
\begin{equation}
\label{2pt}
\chi_R \equiv \frac{ \alpha}{- \beta + \kappa \alpha}
 = \frac{ \alpha_0 + \epsilon \alpha_L + \ldots}
 { -\beta_0 + \kappa \alpha_0 + \epsilon(- \beta_L + \kappa \alpha_L )+ \ldots}
\end{equation}

Close to the critical point where $\kappa \approx \kappa_c$
the leading correction comes only from  $-\beta_L + \kappa_c \alpha_L$
which we intend to compute now.
Consider the normalizable mode $\psi_{\kappa_c}$ at the critical point
defined as:
 \be
 D \psi_{\kappa_c} =0  \,,  \quad  \psi_{\kappa_c}
 = 1 (r-r_0)^{-\delta_-} + 0 (r-r_0)^{-\delta_+}
 \,, \quad \psi_{\kappa_c} = a^+ ( r^{-\Delta_-}  + \kappa_c  r^{-\Delta_+})
 \ee 
 where the last two equations are the boundary conditions
 at the horizon and $AdS_4$ boundary respectively. 
Then if we multiply the first equation in (\ref{towork})
by $\psi_{\kappa_c}$ and integrate by parts twice we find,
\begin{equation}
\label{wronsk}
W\left[ \psi_{\kappa_c} \,; \psi_L + \frac{2\pi T}{\sqrt{3} \mu} r \psi_0' \right]_{r_\star}^\infty
= \int_{r_\star}^\infty \psi_{\kappa_c} X_L \psi_0
\end{equation}
where $W[\psi_A\,; \psi_B] =  f_0 r^2 ( \psi_A \psi_B' -  \psi_B \psi_A'  )$ is the Wronskian.
Also close to the critical point $\Sigma_R$ is small so for the
purposes of computing this correction we can set $\hat{\beta}_0= 0$ such
that $\psi_0 = \hat{\alpha}_0 \psi_{\kappa_c}$. Evaluating the Wronskian
we find the desired result,
\begin{equation}
\frac{(-\beta_L + \kappa_c \alpha_L)_{\hat{\beta}_0 = 0}}{\hat{\alpha}_0}
 = - \frac{ 2 \pi T}{\sqrt{3} \mu} a^+ ( \Delta_+ - \Delta_-) \kappa_c 
 + \frac{1} {a^+ (\Delta_+ - \Delta_-) }  \int^\infty_{r_\star (\rm{reg})} dr  \psi_{\kappa_c} X_L \psi_{\kappa_c}
\end{equation}
where the integral is understood to be regulated. Possible
divergences come only from the horizon and the regulators
actually come from the Wronskian in (\ref{wronsk}) evaluated
as $r \rightarrow r_\star$, when one includes the diverging
terms given in (\ref{lead}). 

Finally the leading contribution is,
\be
\frac{-\beta_0 + \kappa \alpha_0}{\hat{\alpha}_0} = 
a_+ (\kappa- \kappa_c) - \frac{ \det L}{ a_+} \Sigma_R
\ee
Compiling all of these results and using (\ref{2pt}) we
arrive at the result quoted in (\ref{greenfull}) where
the constants in $\mathbb{X}$ are given by,
\bea
\label{ccs}
c_T &=& \frac{ (a^+)^2}{ \det L} \frac{ 2 \pi }{  \sqrt{3} \mu}  (\Delta_+ - \Delta_-)  \\
c_p &=&  \left< g^{xx} \right>  \,, \quad
c_\omega =  \left< |g^{tt}| \right> \,, \quad c_q = \left< A_t |g^{tt}| \right> 
\eea
where the angle brackets define the following:
\begin{equation}
\left< Y \right> \equiv \frac{1}{\det L (\Delta_+ - \Delta_-)}
\int_{r_\star (reg)}^\infty \sqrt{-g} \psi_{\kappa_c} Y \psi_{\kappa_c} 
\end{equation}
and where the metric components above are all for the extremal
black hole metric. Note that these integrals are not always divergent,
depending on the value of $\delta_-$. When they are not divergent
since all the integrands are positive it
follows that the constants in (\ref{ccs}) are positive.
This is always the case for $c_p$, it is only the
case for $c_\omega$ when $\delta_- < -1/2$ and it
is only the case for $c_q$ when $\delta_- < 0$.

\sect{ Complete $AdS_4$ expansion and boundary terms}\label{ap:FullAdS4}

We give here the  asymptotic solution to (\ref{eq:1} - \ref{eq:4}) with $V = -6 + m^2 \psi^2 + (\lambda/2) \psi^4$. The expansion will be complete up to and including $\mathcal{O}(r^{-3})$
assuming that $3/2 < \Delta_+ < 3$. To this order, it suffices to take $G=1$ and $ J = \psi^2$.
The result is only a slight generalization of expressions that
can be found in \cite{Amsel:2006uf}.  The generalization consists of 
 allowing for a nonzero chemical potential (we also work in a different coordinate system.)
These are then used to compute the on shell action and boundary terms
that are needed in Section 4.2.

\begin{eqnarray}
\nonumber
\psi &=& \alpha\, r^{- \Delta_- } \left( 1 - \frac{ q^2 \mu^2}{2 (2 \Delta_- -1 )} r^{-2}  \right) + \beta\, r^{-\Delta_+} \left( 1 -   \frac{ q^2 \mu^2}{2 (2 \Delta_+ -1 )} r^{-2} \right)
\\  & & \qquad  + \frac{ \alpha^3 c_\lambda^1 }{ (3 \Delta_- -4)} r^{- 3 \Delta_-}  + \ldots \nonumber   \\
\phi &=& \mu  - \rho  r^{-1} + \frac{ q^2 \mu \alpha^2}{\Delta_- (2 \Delta_- -1 )} r^{ - 2 \Delta_-}
 + \ldots  \nonumber \\
h r^{-1} &=&   1 - \frac{ \alpha \beta \Delta_- \Delta_+}{6 } r^{-3} - \frac{ \alpha^2 \Delta_- }
{4 (2\Delta_- - 1)} r^{-2\Delta_-} + \frac{ \alpha^4  c_\lambda^2}{(3 \Delta_- - 4)} r^{-4\Delta_-} 
 +  \ldots  ,  \qquad \nonumber \\
f h^{-2} &=&  1  -  (m_0/2 ) r^{-3}  +  \ldots 
\label{expimp}
\end{eqnarray}
where the two constants are:
\begin{equation}
c_\lambda^1 = \frac{ \lambda}{2 \Delta_- } + \frac{ \Delta_- ( 5 \Delta_- -3 )}{4 ( 2 \Delta_- -1)},
\qquad
c_\lambda^2 = - \frac{ 3 \lambda}{ 8 (4 \Delta_- -1)} - \frac{  \Delta_-^2 (  26 \Delta_- -15)}
{32 ( 2 \Delta_- -1 ) ( 4\Delta_- -1)}
\end{equation}

Following \cite{Hartnoll:2008kx} the on shell Euclidean action can be shown
to be a total derivative. The boundary terms give:
\begin{equation}
\label{os}
 S_E = \beta V \left( \left. -2 h h' f \right|^{r= \infty} \right)
\end{equation}
where
$V$ is the volume of space and $\beta$ is the inverse temperature. The horizon
does not give any contribution since $f$ vanishes on the horizon. 

Following \cite{Hartnoll:2008kx} the boundary terms needed in Section 4 are (they are all located
at the boundary of $AdS_4$):
\begin{eqnarray}
\label{ct}
S_{ct}^\beta &=& \int d^3 x\left. \sqrt{-g_\infty}  \left( - 2 K +  A_\mu n_\nu F^{\mu \nu}  - 2 \psi n^\mu
\partial_\mu \psi + \Lambda_B + m_B \psi^2 + \lambda_B \psi^4 \right) \right|^{r =\infty} \\
 &=& \beta V \left. h^2 \left( - f' - 4 h' h^{-1} f +  \phi' \phi - 2 f \psi' \psi
+ \sqrt{f} ( \Lambda_B + m_B \psi^2 + \lambda_B \psi^4 ) \right) \right|^{r=\infty} 
\end{eqnarray}
where $g_\infty$ is the metric on a fixed $r$ slice
as $r\rightarrow \infty$. $K$ is the trace of the extrinsic curvature
and $n^\mu$ is the outward pointing unit normal to the boundary.
By demanding that the $S_E + S_{ct}^\beta$ is finite
one can show that the constants in (\ref{ct}) are,
\begin{equation}
\Lambda_B = 4, \qquad m_B = - \Delta_-, \qquad \lambda_B =  -  \frac{3 ( 3 \Delta_-^2 + 4 \lambda)}
{8 ( 4 \Delta_- - 3 ) }
\end{equation}
Note that the $\psi^4$ term in \ref{ct} is only required when 
$\Delta_- < 3/4$ exactly when $\mathcal{O}^4$ is relevant
in alternative quantization. Plugging (\ref{expimp}) into (\ref{os}) and (\ref{ct})
one finds that the free energy is given by (\ref{free}).



\begin{thebibliography}{99}

\bibitem{Hartnoll:2009sz}
  S.~A.~Hartnoll,
  ``Lectures on holographic methods for condensed matter physics,''
  Class.\ Quant.\ Grav.\  {\bf 26} (2009) 224002
  [arXiv:0903.3246 [hep-th]].
  
\bibitem{McGreevy:2009xe}
  J.~McGreevy,
  ``Holographic duality with a view toward many-body physics,''
  arXiv:0909.0518 [hep-th].
  
\bibitem{Sachdev:2010ch}
  S.~Sachdev,
  ``Condensed matter and AdS/CFT,''
  arXiv:1002.2947 [hep-th].
  
\bibitem{Gubser:2008px}
  S.~S.~Gubser,
  ``Breaking an Abelian gauge symmetry near a black hole horizon,''
  Phys.\ Rev.\  D {\bf 78}, 065034 (2008)
  [arXiv:0801.2977 [hep-th]].
  
\bibitem{Hartnoll:2008vx}
  S.~A.~Hartnoll, C.~P.~Herzog and G.~T.~Horowitz,
  ``Building a Holographic Superconductor,''
  Phys.\ Rev.\ Lett.\  {\bf 101} (2008) 031601
  [arXiv:0803.3295 [hep-th]].
  
\bibitem{Hartnoll:2008kx}
  S.~A.~Hartnoll, C.~P.~Herzog and G.~T.~Horowitz,
  ``Holographic Superconductors,''
  JHEP {\bf 0812} (2008) 015
  [arXiv:0810.1563 [hep-th]].
  
\bibitem{Vecchi:2010dd}
  L.~Vecchi,
  ``Multitrace deformations, Gamow states, and Stability of AdS/CFT,''
  arXiv:1005.4921 [hep-th].
  
\bibitem{Lee:2008xf}
S.~S.~Lee,
``A Non-Fermi Liquid from a Charged Black Hole: a Critical Fermi Ball,''
Phys.\ Rev.\  D {\bf 79} (2009) 086006
[arXiv:0809.3402 [hep-th]].

\bibitem{Liu:2009dm}
  H.~Liu, J.~McGreevy and D.~Vegh,
  ``Non-Fermi liquids from holography,''
  arXiv:0903.2477 [hep-th].


\bibitem{Cubrovic:2009ye}
M.~Cubrovic, J.~Zaanen and K.~Schalm,
``String Theory, Quantum Phase Transitions and the Emergent Fermi-Liquid,''
Science {\bf 325} (2009) 439
[arXiv:0904.1993 [hep-th]].

\bibitem{Faulkner:2009wj}
  T.~Faulkner, H.~Liu, J.~McGreevy and D.~Vegh,
  ``Emergent quantum criticality, Fermi surfaces, and AdS2,''
  arXiv:0907.2694 [hep-th].


\bibitem{Faulkner:2010da}
T.~Faulkner, N.~Iqbal, H.~Liu, J.~McGreevy and D.~Vegh,
``From Black Holes to Strange Metals,''
arXiv:1003.1728 [hep-th].


  
 
\bibitem{Denef:2009tp}
  F.~Denef and S.~A.~Hartnoll,
  ``Landscape of superconducting membranes,''
  Phys.\ Rev.\  D {\bf 79} (2009) 126008
  [arXiv:0901.1160 [hep-th]].
  
  \bibitem{BF82}
	P. Breitenlohner and D. Z. Freedman, ``Positive energy in Anti-de Sitter backgrounds and gauged extended supergravity,Ó Phys. Lett. \textbf{B115} (1982)
197. 



\bibitem{Witten:2001ua}
  E.~Witten,
  ``Multi-trace operators, boundary conditions, and AdS/CFT correspondence,''
  arXiv:hep-th/0112258.
  
\bibitem{Berkooz:2002ug}
  M.~Berkooz, A.~Sever and A.~Shomer,
  ``Double-trace deformations, boundary conditions and spacetime
  singularities,''
  JHEP {\bf 0205} (2002) 034
  [arXiv:hep-th/0112264];
  A.~Sever and A.~Shomer,
  ``A note on multi-trace deformations and AdS/CFT,''
  JHEP {\bf 0207} (2002) 027
  [arXiv:hep-th/0203168].
  
\bibitem{Klebanov:1999tb}
  I.~R.~Klebanov and E.~Witten,
  ``AdS/CFT correspondence and symmetry breaking,''
  Nucl.\ Phys.\  B {\bf 556} (1999) 89
  [arXiv:hep-th/9905104].

  
   
 
  
\bibitem{Faulkner:2010fh}
  T.~Faulkner, G.~T.~Horowitz and M.~M.~Roberts,
  ``New stability results for Einstein scalar gravity,''
  arXiv:1006.2387 [hep-th].
  
\bibitem{Iqbal:2010eh}
N.~Iqbal, H.~Liu, M.~Mezei and Q.~Si,
``Quantum Phase Transitions in Holographic Models of Magnetism and   Superconductors,''
arXiv:1003.0010 [hep-th].
  
 
\bibitem{Faulkner:2010tq}
  T.~Faulkner and J.~Polchinski,
  ``Semi-Holographic Fermi Liquids,''
  arXiv:1001.5049 [hep-th].
  
  
  \bibitem{hertz}
J. A. Hertz, ``Quantum critical phenomena,''
Phys. Rev. B 14, 1165 (1976);\\
 T. Moriya, {\it Spin
Fluctuations in Itinerant Electron Magnetism,} Springer-
Verlag, Berlin (1985);

\bibitem{millis}
 A. J. Millis, ``Effect of a nonzero temperature on quantum critical points in itinerant fermions systems,'' Phys. Rev. B, 48, 7183
(1993).

\bibitem{vojtarev}
H. v. L\"{o}hneysen, A. Rosch, M. Vojta, and P. W\"{o}lße, 
``Fermi-Liquid Instabilities at Magnetic Quantum Phase Transitions'' 
Rev. Mod. Phys. 79, 1015Ð1075 (2007)


\bibitem{sirev}
P.~Gegenwart and Q.~Si and F.~Steglich,
``Quantum criticality in heavy-fermion metals''
Nat. Phys. 4 186-197, (2008)  [arXiv:0712.2045]

\bibitem{si1}
Q. Si, S. Rabello, K. Ingersent, and J. L. Smith, 
``Locally critical quantum phase transitions in strongly correlated 
metals'',
Nature  413, 804Ð808 (2001). 
[arXiv:0903.3246 [hep-th]]. 

\bibitem{si2}
Q. Si, S. Rabello, K. Ingersent, and J. L. Smith, 
``Local fluctuations in quantum critical metals'', Phys. Rev. B 68, 115103 
(2003). [arXiv:cond-mat/0202414]

\bibitem{senthil1}
T. Senthil, S. Sachdev, and M. Vojta, ``Fractionalized Fermi liquids'',
Phys. Rev. Lett. 90, 216403 (2003). 

\bibitem{senthil2}
T. Senthil, M. Vojta, and S. Sachdev,
``Weak magnetism and non-Fermi liquids near heavy-fermion critical points'', 
 Phys. Rev. B 69, 035111 (2004). 

\bibitem{siedmft}
Smith, J. L., and Q. Si, ``Spatial Correlations in Dynamical Mean Field Theory'',
2000, Phys. Rev. B 61, 5184. [arXiv:cond-mat/9903083]



\bibitem{Aprile:2010yb}
  F.~Aprile, S.~Franco, D.~Rodriguez-Gomez and J.~G.~Russo,
  ``Phenomenological Models of Holographic Superconductors and Hall currents,''
  JHEP {\bf 1005}, 102 (2010)
  [arXiv:1003.4487 [hep-th]].
  
\bibitem{Gubser:2002vv}
  S.~S.~Gubser and I.~R.~Klebanov,
  ``A universal result on central charges in the presence of double-trace
  deformations,''
  Nucl.\ Phys.\  B {\bf 656} (2003) 23
  [arXiv:hep-th/0212138].
  
\bibitem{Ishibashi:2004wx}
  A.~Ishibashi and R.~M.~Wald,
  ``Dynamics in non-globally hyperbolic static spacetimes. III: anti-de  Sitter
  spacetime,''
  Class.\ Quant.\ Grav.\  {\bf 21}, 2981 (2004)
  [arXiv:hep-th/0402184].
  
\bibitem{Hertog:2005hu}
  T.~Hertog and G.~T.~Horowitz,
  ``Holographic description of AdS cosmologies,''
  JHEP {\bf 0504}, 005 (2005)
  [arXiv:hep-th/0503071].
  
  \bibitem{Gauntlett:2009dn}
  J.~P.~Gauntlett, J.~Sonner and T.~Wiseman,
  ``Holographic superconductivity in M-Theory,''
  Phys.\ Rev.\ Lett.\  {\bf 103} (2009) 151601
  [arXiv:0907.3796 [hep-th]].
  
\bibitem{Gubser:2009gp}
  S.~S.~Gubser, S.~S.~Pufu and F.~D.~Rocha,
  ``Quantum critical superconductors in string theory and M-theory,''
  Phys.\ Lett.\  B {\bf 683}, 201 (2010)
  [arXiv:0908.0011 [hep-th]].

\bibitem{Cadoni:2009xm}
  M.~Cadoni, G.~D'Appollonio and P.~Pani,
  ``Phase transitions between Reissner-Nordstrom and dilatonic black holes in
  4D AdS spacetime,''
  JHEP {\bf 1003}, 100 (2010)
  [arXiv:0912.3520 [hep-th]].

\bibitem{Son:2002sd}
D.~T.~Son and A.~O.~Starinets,
``Minkowski-Space Correlators in AdS/CFT Correspondence: Recipe and   Applications,''
JHEP {\bf 0209} (2002) 042
[arXiv:hep-th/0205051].


\bibitem{Iqbal:2008by}
N.~Iqbal and H.~Liu,
``Universality of the Hydrodynamic Limit in AdS/CFT and the Membrane   Paradigm,''
Phys.\ Rev.\  D {\bf 79} (2009) 025023
[arXiv:0809.3808 [hep-th]].



\bibitem{exp}
A. Schr\"{o}der, G. Aeppli, R. Coldea, M. Adams, O. Stock- 
ert, H. v. L\"{o}hneysen, E. Bucher, R. Ramazashvili, and 
P. Coleman, Nature 407, 351 (2000). 
``Onset of antiferromagnetism in heavy-fermion metals''

\bibitem{Jensen:2010ga}
K.~Jensen, A.~Karch, D.~T.~Son and E.~G.~Thompson,
``Holographic Berezinskii-Kosterlitz-Thouless Transitions,''
Phys.\ Rev.\ Lett.\  {\bf 105} (2010) 041601
[arXiv:1002.3159 [hep-th]].

\bibitem{D'Hoker:2010ij}
E.~D'Hoker and P.~Kraus,
``Magnetic Field Induced Quantum Criticality via New Asymptotically $\mathrm{AdS}_5$   Solutions,''
arXiv:1006.2573 [hep-th].




\bibitem{Gubser:2009cg}
  S.~S.~Gubser and A.~Nellore,
  ``Ground states of holographic superconductors,''
  Phys.\ Rev.\  D {\bf 80} (2009) 105007
  [arXiv:0908.1972 [hep-th]].

  
\bibitem{Charmousis:2010zz}
  C.~Charmousis, B.~Gouteraux, B.~S.~Kim, E.~Kiritsis and R.~Meyer,
  ``Effective Holographic Theories for low-temperature condensed matter
  systems,''
  arXiv:1005.4690 [hep-th].
  

\bibitem{Horowitz:2009ij}
  G.~T.~Horowitz and M.~M.~Roberts,
  ``Zero Temperature Limit of Holographic Superconductors,''
  JHEP {\bf 0911} (2009) 015
  [arXiv:0908.3677 [hep-th]].

\bibitem{Balasubramanian:2010uw}
  K.~Balasubramanian and J.~McGreevy,
  ``The particle number in Galilean holography,''
  arXiv:1007.2184 [hep-th].
  
\bibitem{Gubser:2000nd}
  S.~S.~Gubser,
  ``Curvature singularities: The good, the bad, and the naked,''
  Adv.\ Theor.\ Math.\ Phys.\  {\bf 4} (2000) 679
  [arXiv:hep-th/0002160].
  
\bibitem{Faulkner:2008hm}
T.~Faulkner and H.~Liu,
``Condensed Matter Physics of a Strongly Coupled Gauge Theory with Quarks:   Some Novel Features of the Phase Diagram,''
arXiv:0812.4278 [hep-th].

  
\bibitem{Maldacena:1998uz}
  J.~M.~Maldacena, J.~Michelson and A.~Strominger,
  ``Anti-de Sitter fragmentation,''
  JHEP {\bf 9902} (1999) 011
  [arXiv:hep-th/9812073].
  
\bibitem{Sachdev:2010um}
S.~Sachdev,
``Holographic Metals and the Fractionalized Fermi Liquid,''
arXiv:1006.3794 [hep-th].

\bibitem{Si}    S.~J. Yamamoto and Q. Si,
``Global Phase Diagram of the Kondo Lattice: From Heavy Fermion Metals to Kondo Insulators'',
  Journal of Low Temperature Physics {\bf 161}, (2010)
[arXiv:1006.4868]
 
  
\bibitem{Horowitz:2010gk}
  G.~T.~Horowitz,
  ``Introduction to Holographic Superconductors,''
  arXiv:1002.1722 [hep-th].
  
\bibitem{Hartnoll:2007ip}
  S.~A.~Hartnoll and C.~P.~Herzog,
  ``Ohm's Law at strong coupling: S duality and the cyclotron resonance,''
  Phys.\ Rev.\  D {\bf 76} (2007) 106012
  [arXiv:0706.3228 [hep-th]].
  
  
\bibitem{Jensen:2010vx}
K.~Jensen,
``More Holographic Berezinskii-Kosterlitz-Thouless Transitions,''
arXiv:1006.3066 [hep-th].


\bibitem{Evans:2010hi}
N.~Evans, A.~Gebauer, K.~Y.~Kim and M.~Magou,
``Phase Diagram of the D3/D5 System in a Magnetic Field and a Bkt   Transition,''
arXiv:1003.2694 [hep-th].


\bibitem{Pal:2010gj}
S.~S.~Pal,
``Quantum Phase Transition in a Dp-Dq System,''
Phys.\ Rev.\  D {\bf 82} (2010) 086013
[arXiv:1006.2444 [hep-th]].



 
\bibitem{Kaplan:2009kr}
D.~B.~Kaplan, J.~W.~Lee, D.~T.~Son and M.~A.~Stephanov,
``Conformality Lost,''
Phys.\ Rev.\  D {\bf 80} (2009) 125005
[arXiv:0905.4752 [hep-th]].


\bibitem{Denef:2009kn}
F.~Denef, S.~A.~Hartnoll and S.~Sachdev,
``Black Hole Determinants and Quasinormal Modes,''
Class.\ Quant.\ Grav.\  {\bf 27} (2010) 125001
[arXiv:0908.2657 [hep-th]].


\bibitem{Denef:2009yy}
F.~Denef, S.~A.~Hartnoll and S.~Sachdev,
``Quantum Oscillations and Black Hole Ringing,''
Phys.\ Rev.\  D {\bf 80} (2009) 126016
[arXiv:0908.1788 [hep-th]].


\bibitem{Hartnoll:2009kk}
S.~A.~Hartnoll and D.~M.~Hofman,
``Generalized Lifshitz-Kosevich Scaling at Quantum Criticality from the   Holographic Correspondence,''
arXiv:0912.0008 [cond-mat.str-el].



\bibitem{Hartman:2010fk}
T.~Hartman and S.~A.~Hartnoll,
``Cooper Pairing Near Charged Black Holes,''
JHEP {\bf 1006} (2010) 005
[arXiv:1003.1918 [hep-th]].


\bibitem{Anninos:2010sq}
D.~Anninos, S.~A.~Hartnoll and N.~Iqbal,
``Holography and the Coleman-Mermin-Wagner Theorem,''
arXiv:1005.1973 [hep-th].

\bibitem{Hartnoll:2009ns}
S.~A.~Hartnoll, J.~Polchinski, E.~Silverstein and D.~Tong,
``Towards Strange Metallic Holography,''
JHEP {\bf 1004} (2010) 120
[arXiv:0912.1061 [hep-th]].

\bibitem{Kachru:2008yh}
S.~Kachru, X.~Liu and M.~Mulligan,
``Gravity Duals of Lifshitz-Like Fixed Points,''
Phys.\ Rev.\  D {\bf 78} (2008) 106005
[arXiv:0808.1725 [hep-th]].

\bibitem{Goldstein:2009cv}
K.~Goldstein, S.~Kachru, S.~Prakash and S.~P.~Trivedi,
``Holography of Charged Dilaton Black Holes,''
arXiv:0911.3586 [hep-th].







\bibitem{D'Hoker:2010rz}
E.~D'Hoker and P.~Kraus,
``Holographic Metamagnetism, Quantum Criticality, and Crossover Behavior,''
JHEP {\bf 1005} (2010) 083
[arXiv:1003.1302 [hep-th]].


\bibitem{Amsel:2006uf}
  A.~J.~Amsel and D.~Marolf,
  ``Energy bounds in designer gravity,''
  Phys.\ Rev.\  D {\bf 74} (2006) 064006
  [Erratum-ibid.\  D {\bf 75} (2007) 029901]
  [arXiv:hep-th/0605101].



\end{thebibliography}
\end{document}